\documentclass[prb,twocolumn,showpacs,superscriptaddress]{revtex4-1}
\usepackage[english]{babel}
\usepackage[dvips]{graphicx}
\usepackage{color}
\usepackage{latexsym}
\usepackage{amsmath}
\usepackage{amsthm}
\usepackage{amsfonts}
\usepackage{amssymb}
\usepackage{romannum}
\usepackage{braket}
\usepackage[T1]{fontenc}

\begin{document}
\newcommand{\WB}[1]{\textbf{WB: #1}}
\newcommand{\GR}[1]{\textbf{GR: #1}}
\newcommand{\RK}[1]{\textbf{\textcolor{red}{ #1}}}
\newcommand{\NP}[1]{\textcolor{blue}{ #1}}

\newcommand{\sgn}{\mathrm{sgn}}
\newcommand{\tj}{\mathrm{tj}}
\newcommand{\cj}{\mathrm{cj}}
\renewcommand{\i}{\mathrm{i}}
\renewcommand{\d}{\mathrm{d}}
\newcommand{\R}{\mathrm{R}}
\renewcommand{\L}{\mathrm{L}}
\newcommand{\hc}{\mathrm{h.c.}}
\newcommand{\kb}{k_\mathrm{B}}

\pagenumbering{arabic}

\title{Nonequilibrium Andreev bound states population in short superconducting junctions coupled to a resonator}
\author{Raffael L. Klees}
\affiliation{Fachbereich Physik, Universit{\"a}t Konstanz, D-78457 Konstanz, Germany}
\author{Gianluca Rastelli}
\affiliation{Zukunftskolleg, Fachbereich Physik, Universit{\"a}t Konstanz, D-78457, Konstanz, Germany}
\author{Wolfgang Belzig}
\affiliation{Fachbereich Physik, Universit{\"a}t Konstanz, D-78457 Konstanz, Germany}

%
%
%
%
\begin{abstract}
Inspired by recent experiments, we study a short superconducting junction of length $L\ll \xi$ (coherence length) inserted in a dc-SQUID containing an ancillary Josephson tunnel junction. We evaluate the nonequilibrium occupation of the Andreev bound states (ABS) for the case of a conventional junction and a topological junction, with the latter case of ABS corresponding to a Majorana mode. We take  into account small phase fluctuations of the Josephson tunnel junction, acting as a damped  LC resonator,
and analyze the role of the distribution of the quasiparticles of the continuum assuming that 
these quasiparticles are in thermal distribution with an effective temperature different 
from the environmental temperature. We also discuss the effect of strong photon irradiation in the junction leading to a nonequilibrium occupation of the ABS. We systematically compare the occupations of the bound states and the supercurrents carried by these states for conventional and topological junctions.
\end{abstract}

\date{\today}
\maketitle

%
%
%
\section{Introduction}
\label{sec:intro}
Recent experiments investigated superconducting junctions containing  
atomic contacts or semiconductor nanowires (NW) with the objective to realize a 
novel type of versatile superconducting junction beyond the standard 
Josephson tunnel junctions. 
Superconducting atomic contacts (SAC) \cite{Scheer:1997ih,Goffman:2000bl,Scheer:2001ju} 
are the simplest example of a short junction hosting doublets of localized Andreev bound states  (ABS) 
that carry the supercurrent in the junction. 
During the last years, a new class of experiments showed the possibility of driving transitions between these  ABS formed in the SAC \cite{Bretheau:2013by,Bretheau:2013bt,Janvier:2015fw}. 
Most importantly, such an Andreev spectroscopy allows for the detection of the occupation of the ABS.
In semiconductor NW combining hybrid properties as strong spin-orbit interaction and superconducting proximity 
effect, ABS corresponding to Majorana modes 
are expected to emerge when the system is driven in a topological range of parameters
\cite{Kwon:2004gf,Fu:2008gu,Nilsson:2008dj,Law:2009gf,Fu:2009hd,Alicea:2010hy,Stanescu:2010bx,Oreg:2010gk,Flensberg:2010bb,Sau:2010gh,Lutchyn:2011jb,Stanescu:2011bg,Badiane:2013gc,Houzet:2013dl,Affleck:2014,Virtanen:2013ew,Hansen:2016de,SanJose:2012gp,SanJose:2013gv,Cayao:2015ge,SanJose:2014jq,vanHeck:2017ic}
(For  a review about Majoranas we refer 
to [\onlinecite{Alicea:2012-review,Leijnse:2012ez,Beenakker:2013tp,Stanescu:2013cl,Beenakker:2015dr,DasSarma:2015jf,Franz:2013kz}]). 
Experiments confirmed several theoretical predictions as  
the zero-bias conductance peak \cite{Mourik:2012je,Rokhinson:2012ep,Zhang:2016vl,Deng:2016de,Nichele:2017ab} or the fractional ac-Josephson effect \cite{Das:2012hi,Deacon:2017da}.
Other experiments in topological junctions also confirmed characteristic features of highly transmitting conductance channels 
which are compatible with the theoretically predicted topological properties. 
For instance, the edge supercurrent associated with the helical edge states was observed in 
two-dimensional HgTe/HgCdTe quantum wells \cite{Hart:2014fq} and 
evidence of the nonsinusoidal phase-supercurrent relation was also reported 
in other works \cite{Sochnikov:2013dc,Sochnikov:2015hl,Kurter:2015hu,Murani:2017kl}. 

SAC or NW are promising for realizing a new qubit architecture in which the information 
is encoded by microscopic degrees of freedom, i.e. the ABS \cite{Lantz:2002ei,Zazunov:2003jm,Zazunov:2005ec,Chtchelkatchev:2003ji,Padurariu:2010fr}, 
rather than the macroscopic BCS condensate, as in conventional superconducting qubits.
Additionally, for NW, the ballistic regime is now within the reach of the experimental devices \cite{Higginbotham:2015ii,Zhang:2016vl,Albrecht:2017hz}  
and the spectroscopic measurement has been now accomplished\cite{vanWoerkom:2017gl} 
using the same method employed in SAC \cite{Bretheau:2013by}.
Andreev spectroscopy, based on employing the microwave signal,
represents a fundamental tool not only for spectroscopy and characterization. It represents a crucial issue towards the coherent control of the Andreev qubits.
A first experiment has already reported coherent quantum manipulation of ABS
in superconducting atomic contacts \cite{Janvier:2015fw}. 
Moreover, junctions formed on ballistic NW have the advantage of being gate-tunable,  
an intriguing property that was used  to devise a new superconducting qubit, 
i.e. the gatemon \cite{Larsen:2015cp}.  
Finally, superconducting topological junctions based on NW enable 
topological protection against dissipation and decoherence that is based on the different 
fermionic parity between two degenerate ground 
states \cite{Kitaev:2001gb,Sarma:2006ch,DasSarma:2015jf,Aasen:2016bja,Karzig:2017if}.

An important and common problem in superconducting junctions is the nonequilibrium
population of the long-lived continuum quasiparticles (QPs). 
In superconducting junctions, it is well known that the population of these QPs lying in the continuum above the gap is not exponentially 
suppressed at low temperature as expected by assuming thermal equilibrium in the 
system \cite{Aumentado:2004ij,Martinis:2009bd,deVisser:2011cu,Lenander:2011dp,Catelani:2011cf,Sun:2012gl,Riste:2013il,Pop:2014gg}.
The underlying mechanism for their relaxation dynamics and their nonequilibrium properties are not fully understood \cite{Bespalov:2016fk}. 
QP excitations can compromise the performance of superconducting devices, 
causing high-frequency dissipation, decoherence in qubits
and braiding errors in proposed Majorana-based topological qubits \cite{Rainis:2012fx,Yang:2014gb,Ng:2015fva}. 
Previous experiments  reported the observation of nonequilibrium Andreev populations and
relaxation in atomic contacts \cite{Zgirski:2011dx}  by measurements of switching currents 
\cite{DellaRocca:2007iy,Bretheau:2012fg}. 
A nonequilibrium QP population was also reported in Aluminum nanobridges with 
submicron constrictions \cite{LevensonFalk:2014en}. 
However, in superconducting junctions with ballistic semiconductor NW, 
a parity lifetime (poisoning time) of the bound state exceeding 10\,ms was reported \cite{Higginbotham:2015ii}.
Motivated by Andreev spectroscopy experiments in short SACs, 
previous theoretical works tackle the problem of the nonequilibrium occupation of the ABS
\cite{Olivares:2014ig,Zazunov:2014kn,Riwar:2014ih,Riwar:2015cv,Riwar:2015it} and a short topological 
junction \cite{Peng:2016bf}.

\begin{figure}[t!]
	\centering
	\includegraphics[width=0.99\linewidth]{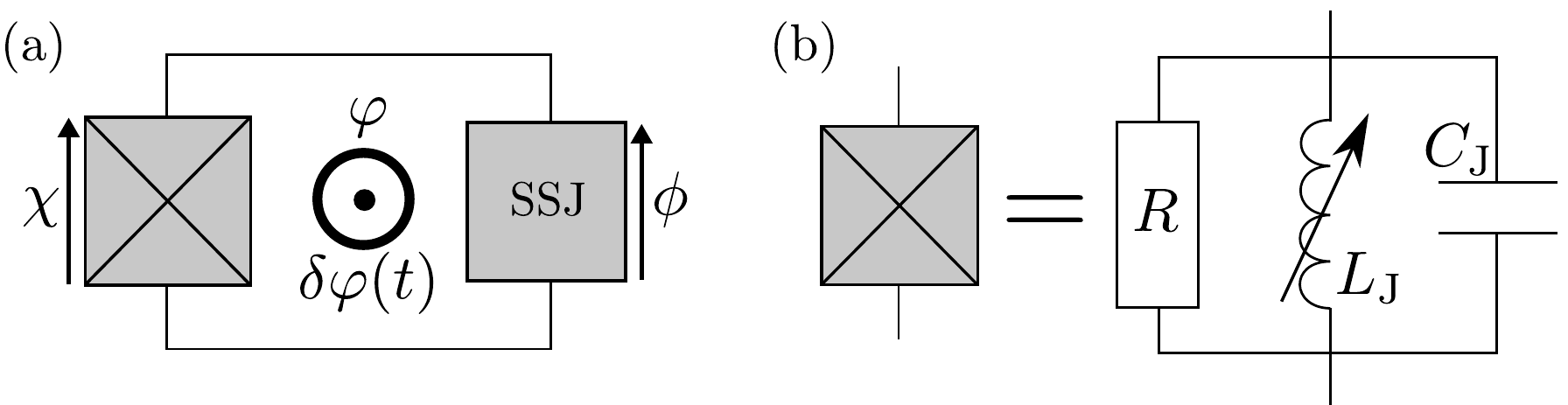}
	\caption{Model of the system. (a) The dc-SQUID is formed by two junctions. The left junction, having a phase difference $\chi$ between the two superconducting leads, is a Josephson tunnel junction with energy $E_\mathrm{J}$. The right junction, having a phase difference $\phi$, is a short superconducting junction (SSJ). The SQUID is penetrated by a magnetic flux $\varphi + \delta\varphi(t)$. (b) For small phase difference fluctuations, the Josephson junction behaves as a damped LC resonator with capacitance $C_\mathrm{J}$ and inductance $L_\mathrm{J} = \Phi_0^2/(4\pi^2 E_\mathrm{J})$. A resistance $R$ accounts for a finite damping on this resonator.}
	\label{fig:model}
\end{figure}

%
%
%
\section{Summary of the main results}
\label{sec:summary}

In this work we discuss the nonequilibrium state of the ABS hosted in a short superconducting junction (SSJ) inserted in a branch of a dc-SQUID, as sketched in Fig.\,\ref{fig:model}(a). 
This system models the experimental setup used in the experiments of the 
Saclay's group\cite{Bretheau:2013by,Bretheau:2013bt,Zgirski:2011dx,DellaRocca:2007iy,Bretheau:2012fg}
and the recent experiment of the Delft's group\cite{vanWoerkom:2017gl}. 
The Josephson junction is in a regime of parameters in which it operates as a damped harmonic LC resonator, as sketched in Fig.\,\ref{fig:model}(b).

\begin{figure}[t!]
	\centering
	\includegraphics[width=0.9\linewidth]{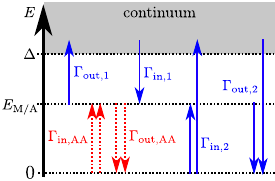}
	\caption{Possible transitions in the SSJ. 
		For both the topological and the conventional junction, there are four transition rates (solid blue arrows) which change the fermionic parity of 
		the ABS in a conventional (Andreev energy $E_\mathrm{A}$) and a topological junction (Andreev energy $E_\mathrm{M}$), respectively. 
		The processes with out-rates $\Gamma_{\mathrm{out,}i}$ empty the bound state, while the
		processes with in-rates $\Gamma_{\mathrm{in,}i}$ fill the bound state. 
		The index $i=1,2$ labels direct transitions between the discrete state and the continuum with or without involving the ground state. 
		For the conventional junction, there are additional parity-conserving processes with rates  $\Gamma_\mathrm{in/out,AA}$
		between the gound state $E = 0$ and the ABS $E_\mathrm{A}$ (dashed red arrows), 
		which are absent for the topological junction since the bound states are nondegenerate.}
	\label{fig:transitions}
\end{figure}

Since the exact microscopic origin of the significant nonequilibrium population of QPs 
in the continuum part of the spectrum is not known, we describe it phenomenologically using 
a thermal distribution but with an effective temperature larger than the bulk temperature of the dc-SQUID.
We show that the steady state nonequilibrium occupation of the ABS is ruled by the microscopic, fundamental fermionic parity changing processes involving the QPs in the continuum\cite{Kos:2013hl,Vayrynen:2015ga} and the emission/absorption of photons with the environment, alias the LC resonator, see  Fig.\,\ref{fig:transitions}.
Although our results rely on a specific choice of electromagnetic environment coupled to the short junction,
we can predict the behavior of the occupations of the ABS as a function of the oscillator's frequency and the phase difference controlled by the dc magnetic flux, see Fig.\,\ref{fig:occupationResonator}.
We found regions in which the occupations are low and close to thermal equilibrium (approximately the ground state) and regions in which we have a population inversion with the ABS of higher energy being almost fully occupied.

We systematically compare these results for the case of a conventional SSJ with the case of topological SSJ hosting Majorana zero modes.
The ABS occupations show remarkable differences between these two cases from which one can infer the peculiar features of the topological junction.  

For both cases, we also discuss the nonequilibrium occupations of the ABS of the SSJ in the presence of an applied ac-drive implemented via ac-modulation of the magnetic flux in the dc-SQUID (that is the phase drop across the junction), see Fig.\,\ref{fig:occupationMicrowaveConstantResonator}.
Then, the population of the ABS is set by the relative strength between the coupling to the LC oscillator and the ac-drive.
The behaviour of the population is determined by the interplay of these two effects. 
In some regions, the ac-drive dominates and restores the thermal occupation of the ABS as expected from the bulk temperature of the system.

Remarkably, all these features appear in the current-phase relation that shows characteristic jumps associated with the switching of the ABS from one regime of the steady state occupation to another one, with the largest jump of the current associated with the switching from the excited to the ground state (or vice versa), see Figs.\,\ref{fig:CurrentPhaseRelationMBS1}, \ref{fig:CurrentPhaseRelationABS1} and \ref{fig:CurrentWithMicrowave}.
These jumps resemble the discontinuities being smoothed out by the thermal spreading
which are theoretically associated with a topological transition at zero temperature \cite{Marra:2016}.
Hence, our study offers another possible mechanism that could explain the observation of the anomalous features eventually observed in the current-phase relation.

This paper is structured as follows: In Sec.\,\ref{sec:description}, we define the total dc-SQUID setup which is sketched in Fig.\,\ref{fig:model} together with its basic properties, followed by Sec.\,\ref{sec:model} with a detailed discussion of the Hamiltonians of the SSJ, the LC resonator and the interactions in the SQUID.
In Sec.\,\ref{sec:dynamics}, we give the rate equations for the bound states in terms of transition rates between discrete and continuum states of the SSJ, see Fig.\,\ref{fig:transitions}. We discuss the stationary solutions of the rate equations together with the resulting supercurrents in Sec.\,\ref{sec:results}.
We summarize our results in Sec.\,\ref{sec:conclusion} with a discussion about the experimental observation of the nonequilibrium occupation of the ABS.

%
%
%
\section{Description of the system}
\label{sec:description}

The dc-SQUID is formed by a Josephson  junction and a short superconducting junction (SSJ), as sketched in Fig.\,\ref{fig:model}(a).
We assume that the Josephson tunnel junction itself is formed by a second, smaller SQUID such 
that it behaves as a single Josephson  junction of tunable Josephson energy $E_\mathrm{J}$.
The phase difference between the two superconducting leads of each junction is described by $\chi$ for the Josephson junction and $\phi$ for the SSJ.
The Josephson junction behaves as a damped LC resonator having a capacitance $C_\mathrm{J}$ and a tunable inductance $L_\mathrm{J} = \Phi_0^2/(4\pi^2 E_\mathrm{J})$, as shown in Fig.\,\ref{fig:model}(b), with the flux quantum $\Phi_0 = \hbar / 2e$. 

The whole dc-SQUID is penetrated by a dc magnetic flux with small ac part $\varphi + \delta\varphi(t)$.
The phase differences in the superconducting ring are linked by
\begin{equation}
	\phi - \chi = \varphi + \delta\varphi(t) \, .
	\label{eqn:link}
\end{equation}

In the limit in which the Josephson energy of the tunnel junction is larger than the superconducting coupling of the SSJ,
the phase difference essentially drops on the SSJ and the external magnetic flux enables control of the phase difference, i.e. $\phi \approx \varphi$, assuming the ac-drive $\delta\varphi(t)$ is small.

In the ballistic regime, we discuss a nanostructure characterized by one conducting transmission channel which gives rise to ABS when it is embedded between two superconducting 
leads \cite{Kulik:1969JETP,Furusaki:1991bb,Beenakker:1991,Beenakker:1991fv}. 
For the case of a conventional junction, we consider also a delta-like barrier in the nonsuperconducting region to mimic the effect of finite transmission\cite{Bagwell:1992gz,Schussler:1993jp}.

Both the case of the short topological and the short conventional junction is described by the model Hamiltonian of the total system given by
\begin{equation}
H = H_\mathrm{SSJ} + H_\mathrm{res} + H_\mathrm{int} + H_\mathrm{mw}(t)\, ,
\label{eqn:totalHamiltonian}
\end{equation}
with the Hamiltonian $H_\mathrm{SSJ}$ of the SSJ, the Hamiltonian of the damped LC resonator $H_\mathrm{res}$ and the interaction $H_\mathrm{int}$ between the SSJ and the resonator. The Hamiltonian $H_\mathrm{mw}(t)$ originates from the ac part of the magnetic flux driving the phase difference at microwave frequencies. Details of the derivation of this model are shown in appendix \ref{appendix:c}.

We assume that the coupling strength between the resonator and the discrete states of SSJ 
is weak enough such that the resonator's damping overwhelms and we can disregard  
the coherent coupling between the resonator and the discrete states of the SSJ.
The coherent coupling leads to an anticrossing in the spectrum \cite{Bretheau:2014bq} and 
it occurs for strong coupling between the ABS in a SAC in high-quality superconducting 
microwave resonators, as observed in Ref.\,[\onlinecite{Janvier:2015fw}].
Such a strong coherent coupling between ABS and high-quality superconducting microwave resonators was proposed as a new architecture for the circuit QED\cite{Bretheau:2014bq,Romero:2012cp} and for photon measurements of the cavity response for the detection of the topological properties of the junction \cite{Dmytruk:2015ez, Dmytruk:2016fj}.
Finally, we treat the microwave source as an incoherent emission or absorption of photons at frequency $\Omega$ in the junction. 
This is valid if the energy $\hbar \Omega$, with the reduced Planck constant $\hbar$, is far away from the internal resonance $\Delta E = 2 |E_\mathrm{M,A}|$ 
and for strongly damped QPs in the continuum\cite{Virtanen:2013ew,Bergeret:2010bw,Bergeret:2011gs}\textsuperscript{,}\footnote{The exact treatment of the time-dependent term leading to a coherent evolution 
was studied in previous works, Refs.\,[\onlinecite{Virtanen:2013ew,Bergeret:2010bw,Bergeret:2011gs}] (but neglecting the QPs in the continuum) 
and is beyond the goal of this work. }.
In the rest of the paper, we refer to the Andreev bound states formed in the short conventional junction as ABS, having an energy $E_\mathrm{A}$, 
whereas we refer to the Andreev bound states formed in the short topological junction as Majorana bound states (MBS), having an energy $E_\mathrm{M}$.

%
%
%
\section{Model Hamiltonians}
\label{sec:model}
In this section, we will present the Hamiltonians of the system appearing in Eq.\,\eqref{eqn:totalHamiltonian} which follow from the derivation shown in appendix \ref{appendix:c}. First, in Sec.\,\ref{subsec:ssj}, we give the detailed discussion of the Hamiltonians of the SSJ for the case of a topological (Sec.\,\ref{subsubsec:topological}) or a conventional (Sec.\,\ref{subsubsec:conventional}) junction. We solve the time-independent Bogoliubov-de Gennes (BdG) equations for each case to obtain the full eigensystem of the corresponding junctions. 
The discussion of the Hamiltonian for the large Josephson junction, acting as a damped LC resonator, is provided in Sec.\,\ref{subsec:resonator}. 
Finally, in Sec.\,\ref{subsec:interaction}, we give the explicit expressions for the interaction between the SSJ and the damped LC resonator and the SSJ with the ac microwave field, respectively. Both the resonator and the microwave field couple to the SSJ via the operator of the current through the SSJ. The matrix elements of the current operator are calculated using the eigenstates of the time-independent BdG equations. These matrix elements enter the transition rates between different states of the SSJ which are discussed in detail in Sec.\,\ref{sec:dynamics}.

\subsection{$\mathbf{H_\mathrm{\mathbf{SSJ}}}$ for the short superconducting junction}
\label{subsec:ssj}
\begin{figure}[t]
	\centering
	\includegraphics[width=0.9\linewidth]{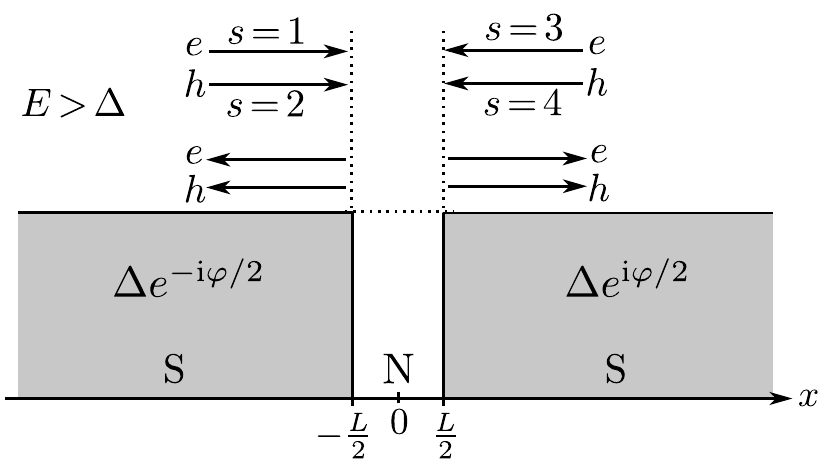}
	\caption{Sketch of the SSJ of length $L$ formed by two superconductors (S) separated by a small normal (N) region. In S, there is an excitation gap of $2\Delta$ around the Fermi energy $E=0$ and there is a superconducting phase difference of $\varphi$ across the junction which is controlled by the dc magnetic flux. For energies $E > \Delta$, we have propagating QPs whose wave functions are obtained by calculation of $s$  scattering states ($s = 1,2,3,4$) labeling the incident QP (see main text). In general, each incident QP produces four outgoing QPs due to normal or Andreev reflection.}
	\label{fig:topJuncCH4}
\end{figure}

The Hamiltonian of the SSJ is given by $H_\mathrm{SSJ}^{(\beta)} = \int \d x \Psi^\dag_{\beta}(x) \mathcal H_{\beta}(x) \Psi_{\beta}^{\phantom{\dag}}(x) /2$ for the topological ($\beta = \tj$) and the conventional ($\beta = \cj$) junction, respectively, with the corresponding BdG Hamiltonian $\mathcal H_{\beta}(x)$ and the Nambu spinor $\Psi_{\beta}^{\phantom{\dag}}(x)$ which will be specified below. The model of the SSJ is sketched in Fig.\,\ref{fig:topJuncCH4}.
The SSJ is assumed to be short, i.e. $\hbar v_\beta / \Delta \gg L$, where $L$ is the junction's length (i.e. the length of the nonsuperconducting region N), $\Delta$ is the absolute value of the pairing potential in the superconducting region S and $v_\beta$ is the Fermi velocity of the SSJ in the topological or conventional case. In the short junction limit, we can consider $L \to 0$ and, therefore, the inhomogeneous gap potential $\Delta(x)$ and the pairing phase $\phi(x)$ become
\begin{align}
\Delta(x) 
= 
\left\{\begin{array}{cc} 
\Delta, & x \neq 0 \\ 0, & x = 0 
\end{array} \right. 
\, , \quad 
\phi(x) 
= \frac{\varphi}{2} \, \sgn(x)  \, , 
\label{ch4:phaseTopJunc}
\end{align}
with the sign function $\sgn(x)$ and $\varphi$ being the total phase difference between the two superconducting leads controlled by the dc part of the magnetic flux. 

We diagonalize the Hamiltonian by solving the corresponding time-independent BdG equations $\mathcal{H}_\beta \Phi_{\beta,n} = E_{\beta,n}  \Phi_{\beta,n}$ with a scattering ansatz to obtain the energy spectrum and the eigenstates. We obtain a discrete spectrum ($n = \pm$) for states bound to the junction with energies $|E| < \Delta$ and a continuous spectrum of scattering states ($n = (E,s)$) at energies $|E| > \Delta$. Here, $s$ labels the four possible incident QPs, with $s = 1$ ($s = 2$) describing an electron-like (hole-like) QP impinging from the left and $s=3$ ($s=4$) describing an electron-like (hole-like) QP impinging from the right lead (see Fig.\,\ref{fig:topJuncCH4}). 
The full expressions of the wave functions $\Phi_{\beta,n}$ are provided in appendix \ref{appendix:a} for the short topological (appendix \ref{subappendix:a1}) and short conventional (appendix \ref{subappendix:a2}) junction. 
These solutions allow us to expand the field operators as $\Psi_{\beta}(x) = \sum_{n} \Phi_{\beta,n}(x) \, \gamma_{\beta,n}$ in terms of these wave functions by introducing fermionic BdG QP operators $\gamma_{\beta,n}$.
Eventually, the diagonalized Hamiltonian of the SSJ is then given by
\begin{equation}
	H_\mathrm{SSJ}^{(\beta)} = \sum_{n} E_{\beta,n} \, \gamma_{\beta,n}^\dag \gamma_{\beta,n}^{\phantom{\dag}} \, .
\end{equation}

\setcounter{paragraph}{0}
\subsubsection{Short topological superconducting junction}
\label{subsubsec:topological}
We model the short topological junction by using the Fu-Kane-Model \cite{Fu:2008gu} of superconductivity-proximized helical edge states in a two-dimensional topological insulator (TI), for which the BdG Hamiltonian is given by
\begin{equation}
\mathcal H_\tj(x) = - \i \hbar v_\tj \sigma_3 \tau_3 \partial_x - \mu \tau_3 + \Delta(x) \, e^{\i\phi(x)\tau_3}  \tau_1 \, , \label{ch4:bdgHamiltonianTopologicalJunction}
\end{equation}
with $\mu$ being the chemical potential and $v_\text{tj}$ being the Fermi velocity of the edge states. 
This Hamiltonian can be related directly to the low-energy Hamiltonian of a spin-orbit coupled NW. As shown in the supplemental material of Ref.\,[\onlinecite{Vayrynen:2015ga}], the low-energy Hamiltonian of a clean NW junction in the limit of a strong magnetic field is the same as for a reflectionless S-TI-S junction, with different velocity $v_\tj$ and pairing gap $\Delta$. In this model, spin and momentum are not independent of each other leading to effectively spinless superconductivity which is necessary for topological superconductivity. This guarantees the existence of a Majorana zero mode.

The matrices $\tau_i$ and $\sigma_i$ are Pauli matrices acting on particle-hole and right/left-movers subspace (which corresponds to spin-space since spin is locked to momentum due to helicity), respectively, of the Nambu space defined by the spinor $\Psi_{\tj}(x) = ( \psi_\uparrow^{\phantom{\dag}}(x) ,  \psi_\downarrow^\dag(x) , \psi_\downarrow^{\phantom{\dag}}(x) , -\psi_\uparrow^\dag(x))^\text{T}$, with the annihilation (creation) operator $\psi_\sigma^{(\dag)}(x)$ of a QP with spin $\sigma$. 
Matrices of different subspaces commute, i.e. $\left[ \tau_i , \sigma_j \right] = 0$. 
Particle-hole symmetry is expressed by the operator $\mathcal S_\tj =  \sigma_2 \tau_2 K$, $K$ meaning complex conjugation, fulfilling $\{\mathcal S_\tj, \mathcal H_\tj(x)\} = 0$.
Diagonalization of the Hamiltonian reveals a single pair of nondegenerate bound states $E_\pm(\varphi) = \pm E_\mathrm{M}(\varphi)$ with the $4\pi$-periodic energy \cite{Alicea:2012-review,Leijnse:2012ez,Beenakker:2013tp,Stanescu:2013cl,Beenakker:2015dr,DasSarma:2015jf} 
\begin{equation}
	E_\mathrm{M}(\varphi) = \Delta \cos\frac{\varphi}{2}
\end{equation}
of the MBS, with the topologically protected zero-energy crossing at $\varphi = \pi$. The current through the topological junction can be expressed as
\begin{equation}
	I_\mathrm{M}(\varphi) = \frac{1}{\Phi_0} \frac{\partial E_\mathrm{M}}{\partial \varphi} \left( n_\mathrm{M} - \frac{1}{2}\right) \, ,
	\label{eqn:MBScurrent}
\end{equation}
with  the occupation $n_\mathrm{M} \in \{0,1\}$ of the MBS.

\subsubsection{Short conventional superconducting junction}
\label{subsubsec:conventional}
For the conventional junction, we start from a general second-quantized density Hamiltonian and linearize around the Fermi surface by replacing the fermionic annihilation (creation) operators $\psi_\sigma^{(\dag)}(x)$ of electrons having spin $\sigma =\, \uparrow,\downarrow$ with
\begin{align}
	\psi_\sigma^{(\dag)}(x) = e^{ \pm \i k_\cj x} \psi_{\R\sigma}^{(\dag)}(x) + e^{ \mp \i k_\cj x} \psi_{\L\sigma}^{(\dag)}(x) \, .
\end{align}
By using the Nambu notation, we obtain that the BdG Hamiltonian of spin-up QPs in the short conventional junction takes the form\cite{nabl09}
\begin{multline}
\mathcal H_\cj(x) = - \i \hbar v_\cj  \sigma_3 \tau_3 \partial_x 
\\
+ \hbar v_\cj  Z \, \delta(x) \, \sigma_1 \tau_3 + \Delta(x) \, e^{\i\phi(x)\tau_3}  \tau_1 \, , \label{ch4:bdgHamiltonianConventionalJunction}
\end{multline}
with $v_\text{cj} = \hbar k_\cj / m$ being the Fermi velocity in the conventional junction, the mass $m$ of an electron, $k_\cj$ is the Fermi wave number.

In Eq.\,(\ref{ch4:bdgHamiltonianConventionalJunction}), we model an arbitrary transmission $0 < \mathcal T < 1$ through the conventional junction by including a finite $\delta$-barrier of strength $Z > 0$ at $x=0$ leading to scattering at the interface, turning right- into left-movers and vice versa. We assume the transmission probability $\mathcal T$ to be energy-independent and to be related to the barrier strength $Z$ by the relation $\mathcal T=\cosh^{-2}(Z)\,$ \cite{Kulik:1969JETP,Furusaki:1991bb}.

The matrices $\tau_i$ and $\sigma_i$ are Pauli matrices acting on particle-hole and right/left-mover subspace, respectively, of the Nambu space defined by the spinor $\Psi_{\cj}(x) = \bigl( \psi_{\R\uparrow}^{\phantom{\dag}}(x) , \psi_{\L\downarrow}^{\dag }(x) , \psi_{\L\uparrow}^{\phantom{\dag}}(x) , \psi_{\R\downarrow}^\dag(x)\bigr)^\text{T}$, with the creation (annihilation) operator $\psi_{\alpha\sigma}^{(\dag)}(x)$ of a QP with spin $\sigma$ moving in the direction $\alpha$. 
Again, matrices of different subspaces commute, i.e. $\left[ \tau_i , \sigma_j \right] = 0$. 
For the conventional junction, particle-hole symmetry is described by the operator $\mathcal S_\cj = \i \sigma_1 \tau_2 K$ which fulfills $\{\mathcal S_\cj, \mathcal H_\cj(x)\} = 0$.
Diagonalization of the Hamiltonian reveals a single pair of twofold degenerate 
bound states $E_\pm(\varphi,\mathcal T) = \pm E_\mathrm{A}(\varphi,\mathcal T)$ 
with the $2\pi$-periodic ABS energy \cite{Beenakker:1991fv}
\begin{equation}
E_\mathrm{A}(\varphi,\mathcal T) = \Delta \sqrt{ 1- \mathcal T \sin^2\frac{\varphi}{2} }\, .
\label{eq:ABS}
\end{equation}
For any nonperfect transmission $\mathcal T < 1$ we find $E_\pm(\varphi,\mathcal T) \neq 0$ for all phases $\varphi$, so the spectrum is gapped in this case. The current through the conventional junction can be expressed as
\begin{equation}
I_\mathrm{A}(\varphi,\mathcal T) = \frac{1}{\Phi_0} \frac{\partial E_\mathrm{A}}{\partial \varphi} \left( n_\mathrm{A} - 1 \right)
\label{eqn:ABScurrent} 
\end{equation}
and it depends on the occupation $n_\mathrm{A} \in \{0,1,2\}$ of the spin-degenerate ABS. In contrast to the short topological junction, the conventional junction has a state of zero current corresponding to $n_\mathrm{A} = 1$.
Finally, the spin-down QPs are described by the same Hamiltonian, given in Eq.\,(\ref{ch4:bdgHamiltonianConventionalJunction}), but with a different spinor given by $\mathcal S_\cj \Psi_{\cj}$.

Comparing the BdG Hamiltonians for the topological and the conventional junction, i.e. Eqs.\,(\ref{ch4:bdgHamiltonianTopologicalJunction}) and (\ref{ch4:bdgHamiltonianConventionalJunction}), we see that they have a similar structure which is due to the definitions of the Nambu spinors $\Psi_{\beta}(x)$. For the case of a conventional junction, we introduce a finite transmission $0 < \mathcal T < 1$ through the junction leading to a finite gap in the Andreev spectrum, see Eq.\,(\ref{eq:ABS}). This is in contrast to the topological case in which the level crossing of the MBS is protected and cannot be removed by a finite transmission. Additionaly, the Hamiltonian of the conventional junction provides states which are not helical, i.e. momentum and spin are indepedent degrees of freedom leading to spin degeneracy of the states.

\subsection{Josephson junction as a dissipative resonator}
\label{subsec:resonator}
We now specify the Hamiltonian $H_\mathrm{res}$ of the damped resonator. 
We assume that the Josephson junction is in the Josephson regime in which the Josephson energy $E_\mathrm{J}$ is large compared to the charging energy $E_\mathrm{C} = (2e)^2/2C_\mathrm{J}$, i.e. $E_\mathrm{J} \gg E_\mathrm{C}$, where $e$ is the elementary charge and $L_\mathrm{J}$ ($C_\mathrm{J}$) is the inductance (capacitance) of the Josephson junction. Since fluctuations in the phase difference $\chi$ are small in this regime, the Josephson junction behaves like a LC resonator (cf. Fig.\,\ref{fig:model}(b)) with an effective Hamiltonian
\begin{equation}
	H_\mathrm{res} 	= \hbar \omega_0 \, b_0^\dag b_0^{\phantom{\dag}}  + H_\mathrm{bath}\, ,
	\label{eqn:HamiltonianResonator}
\end{equation}
where we introduced bosonic creation and annihilation operators $b_0^\dag$ and $b_0^{\phantom{\dag}}$, respectively, together with the Josephson plasma frequency $\omega_0  = \sqrt{2 E_\mathrm{J}  E_\mathrm{C}} / \hbar$.
$H_\mathrm{bath}$ describes the unavoidable dissipation of the LC resonator which is taken into account by assuming a resistor $R$ connected in parallel to the Josephson junction (cf. Fig.\,\ref{fig:model}(b)). 
The bath can be formally described with the Caldeira-Leggett model \cite{Caldeira:1983wy}, 
i.e. coupling the resonator to an infinite set of independent harmonic oscillators producing an Ohmic damping $\gamma$.
The resistor is assumed to be at environmental temperature $T_\mathrm{env}$.
The correlator of the damped LC resonator reads
\begin{multline}
C(t) = \Bigl\langle \bigl(b_0^\dag(t) + b_0^{\phantom{\dag}}(t) \bigr) \, \bigl(b_0^\dag + b_0^{\phantom{\dag}} \bigr)  \Bigr\rangle  \\
= \frac{1}{4} \int_0^\infty \!\!\! \d E \, \chi(E) \, \Bigl( n_\mathrm{B}(E) \, e^{\i E t/\hbar} + \bigl(1 + n_\mathrm{B}(E) \bigr) \, e^{-\i E t/\hbar} \Bigr) \, , \label{eqch5:sumOfCorrelationsCalculated}
\end{multline}
with the Bose-Einstein distribution $n_\mathrm{B}(E) = 1/(e^{E / \kb T_\mathrm{env}}-1)$, the Boltzmann constant $\kb$ and the spectral density of the LC resonator
\begin{equation}
	\chi(E) = \frac{8 \hbar \omega_0 \gamma  E / \pi}{(E^2 - (\hbar\omega_0)^2)^2 + 4 \gamma^2 E^2} \, .
	\label{eqn:spectralDensityResonator}
\end{equation}

\subsection{Interaction with the damped resonator and microwave irradiation}
\label{subsec:interaction}
In this part, we discuss the interactions in the dc-SQUID   between the SSJ and 
the damped LC resonator as well as the effect of a time dependent ac flux. This small ac flux induces an ac phase-drive $\delta\varphi(t)$ given by $\delta\varphi(t) = \delta\varphi \, \sin(\Omega t)$ with microwave frequency $\Omega$ and coupling strength $\delta\varphi$.
The interaction Hamiltonian leading to dynamics in the SSJ is given by
\begin{subequations}
	\begin{align}
		H_\mathrm{int}^{(\beta)} &= \lambda \bigl( b_0^\dag + b_0^{\phantom{\dag}} \bigr) \, \Phi_0 \, I_\beta \, , 
		\label{eqn:interactionRES}
		\\
		H_\mathrm{mw}^{(\beta)}(t) &=  \delta\varphi \, \sin(\Omega t) \, \Phi_0 \, I_\beta \, , \label{eqn:interactionMW}
	\end{align}\label{eqn:interaction}%
\end{subequations}
with the coupling to the resonator $\lambda = \sqrt{E_\mathrm{C}/\hbar\omega_0}$ and the current operator of the SSJ given by
\begin{equation}
	I_\beta = \frac{e v_{\beta}}{2} \Psi_{\beta}^\dag (0) \sigma_3 \Psi_{\beta}^{\ }(0) \, ,
	\label{eqn:currentOperatorSSJ}
\end{equation}
evaluated at the interface $x=0$. 
Again, $\beta = \tj$ ($\beta = \cj$) labels the short topological (conventional) junction. 
We note that a time-dependent ac phase bias induces a time-dependent voltage  $V(t)$ according to the Josephson relation
$V(t) = \Phi_0 \, \partial_t \delta\varphi(t)$
which will be neglected since it only leads to a (time-dependent) renormalization of the energy levels and, thus, will not modify the transition rates in our approach to the dynamics with a master equation\cite{Kos:2013hl}.

%
%
%
\section{Rate equation for $\mathbf{n_\mathrm{\mathbf{M}}}$ and $\mathbf{n_\mathrm{\mathbf{A}}}$}
\label{sec:dynamics}

In this section, we describe the nonequilibrium dynamics of the SSJ by using a rate equation. In both cases, topological and conventional SSJ, there is a single pair of bound states at subgap energies $|E| < \Delta$, as described in Sec.\,\ref{subsec:ssj}. Depending on the type of the SSJ, there are several possible transitions between the ground state at $E=0$, the MBS (ABS) $|E_\mathrm{M}|<\Delta$ ($|E_\mathrm{A}|<\Delta$) and the continuum at $|E| > \Delta$. The transition rates between these different states can be obtained by using Fermi's golden rule.

More formally, the rate equation for the occupation of the bound states can be derived by starting from the time-evolution of the density matrix of the total system and, finally, using a Born-Markov approximation\cite{brpe02} and neglecting any coherence in the system described by off-diagonal elements in the density matrices. 
Tracing out the damped resonator yields a reduced density matrix of the SSJ which is approximated as a direct product of subgap part $\alpha = \mathrm{M}$ ($\alpha = \mathrm{A}$), referring to the MBS (ABS) in the topological (conventional) junction, and continuum (c) part, i.e. $\rho_\mathrm{SSJ} = \rho_\alpha \otimes \rho_\mathrm{c}$.
After tracing over the continuum, we obtain a density matrix of the bound states $\rho_\alpha$ from which one calculates rate equations for occupation probabilities $P_i(t)$ of the bound states.
Such an approach was used, for instance, in Ref.\,[\onlinecite{Zazunov:2014kn}].

Since the SSJ obeys particle-hole symmetry as described in Sec.\,\ref{subsec:ssj}, we can restrict the description to energies $E \geq 0$ because creating an excitation at energy $E > 0$ corresponds to destroying a QP at $-E$.
Finally, we assume that continuum QPs relax fast and that they are described by the Fermi-Dirac distribution $f(E) = 1/(e^{E/ \kb T_\mathrm{qp}} + 1)$ with a QP temperature $T_\mathrm{qp}$.

Notice that we assume $T_\mathrm{qp} \neq T_\mathrm{env}$ to mimic the effective, nonequilibrium distribution of the QPs in the continuum, with $\kb T_\mathrm{env} \ll \hbar \omega_0$ ($n_\mathrm{B}(E) \approx 0$).

\subsection{Topological junction}
\label{subsec:masterEqnTJ}
For the topological junction, the first excited state corresponding to the MBS can only be empty ($i=0$) or occupied with one QP ($i=1$). 
Hence, the full rate equation for the probabilities $P_i(t)$ reads
\begin{equation}
	\frac{\d}{\d t} \begin{pmatrix} P_0(t) \\ P_1(t) \end{pmatrix}
	=
	\begin{pmatrix} -\Gamma_\mathrm{in} & \Gamma_\mathrm{out} \\ \Gamma_\mathrm{in} & -\Gamma_\mathrm{out} \end{pmatrix}
	\begin{pmatrix} P_0(t) \\ P_1(t) \end{pmatrix}
	\label{eqn:probabilitiesTJ}
\end{equation}
with the populating in-rate $\Gamma_\mathrm{in} = \Gamma_\mathrm{in,1}^\mathrm{mw} + \Gamma_\mathrm{in,1}^\mathrm{res} + \Gamma_\mathrm{in,2}^\mathrm{mw} + \Gamma_\mathrm{in,2}^\mathrm{res}$ and the depopulating out-rate $\Gamma_\mathrm{out} = \Gamma_\mathrm{out,1}^\mathrm{mw} + \Gamma_\mathrm{out,1}^\mathrm{res} + \Gamma_\mathrm{out,2}^\mathrm{mw} + \Gamma_\mathrm{out,2}^\mathrm{res}$, cf. Fig.\,\ref{fig:transitions}.
The transition matrix elements obtained from the current operator in Eq.\,(\ref{eqn:currentOperatorSSJ}) in the case of a topological SSJ are explicitly shown in appendix \ref{subappendix:b1}. We introduce the quantity
\begin{equation}
	\rho_\tj^\pm(E) = \frac{\sqrt{\Delta^2 - E_\mathrm{M}^2}}{\Delta^2} \, \frac{\sqrt{E^2 - \Delta^2}}{E \pm E_\mathrm{M}} 
\end{equation}
which has the meaning of an effective density of states resulting from the product of the corresponding matrix element of the current operator and the density of states in a superconductor $D(E) = N_\tj \, E \, / \sqrt{E^2 - \Delta^2}$, with $N_\tj = \mathcal L/\pi \hbar v_\mathrm{tj}$ being the density of states in the normal state. $\mathcal L$ defines a length-scale over which the porpagating QP scattering states of the superconducting contact are defined. The rates for microwave radiation read
	\begin{align}
		\Gamma_\mathrm{out,2/in,1}^\mathrm{mw}  &= \frac{(\delta\varphi)^2 \Delta^2}{16 \hbar}\, \rho_\tj^\pm(\hbar \Omega \mp E_\mathrm{M}) \nonumber \\
		&  \quad \times f(\hbar \Omega \mp E_\mathrm{M}) \,  \Theta(\hbar \Omega - (\Delta \pm E_\mathrm{M})) 
		\label{eqn:MicrowaveEmission}
	\end{align}
	for the photon emission and
	\begin{align}
		\Gamma_\mathrm{in,2/out,1}^\mathrm{mw} &=   \frac{(\delta\varphi)^2 \Delta^2}{16 \hbar}\,  \rho_\tj^\pm(\hbar \Omega \mp E_\mathrm{M}) \nonumber  \\
		 &  \times  \bigl(1- f(\hbar \Omega \mp E_\mathrm{M}) \bigr) \,  \Theta(\hbar \Omega - (\Delta \pm E_\mathrm{M})) \label{eqn:MicrowaveAbsorption}
	\end{align}
	for the photon absorption. The rates associated with the emission and absorption of photons in the damped LC resonator read
\begin{subequations}
	\begin{align}
		\Gamma_\mathrm{out,2/in,1}^\mathrm{res} &=  \frac{\lambda^2 \Delta^2}{16 \hbar}  \int_{\Delta}^{\infty} \!\!\!\!  \d E \, \rho_\tj^\pm(E) \, f(E) \nonumber  \\
		& \qquad  \times  \chi(E \pm E_\mathrm{M}) \,  \bigl(1+n_\mathrm{B}(E \pm E_\mathrm{M}) \bigr) \, , \label{eqn:ratesZeroEnvTemp} \\
		\Gamma_\mathrm{in,2/out,1}^\mathrm{res} &=  \frac{\lambda^2 \Delta^2}{16 \hbar} \int_{\Delta}^{\infty} \!\!\!\!  \d E \, \rho_\tj^\pm(E) \,  \bigl(1-f(E) \bigr) \nonumber \\
		& \quad \qquad \quad \times \chi(E \pm E_\mathrm{M}) \, n_\mathrm{B}(E \pm E_\mathrm{M}) \, .
	\end{align}
\end{subequations}
Due to the nondegeneracy and different fermionic parity of the MBS, 
there is no direct transfer of a Cooper pair between the ground state and the first excited state.
For $T_\mathrm{qp} = 0$ implying $f(E) = 0$, our rates for microwave absorption, given in Eq.\,(\ref{eqn:MicrowaveAbsorption}), coincide with the ones reported in Ref.\,[\onlinecite{Peng:2016bf}] expressed in terms of  the admittance 
$Y(\Omega)$ in a short topological junction.
The notation of the transition rates $\Gamma_{j,k}^l$ can be understood by means of Fig.\,\ref{fig:transitions}, with $j \in \{ \mathrm{in,out}\}$ refering to (out-) in-rates (de-) populating the MBS, $k \in \{1,2\}$ refering to the number of QPs which are involved and $ l \in \{ \mathrm{mw,res}\}$ is labeling the source of perturbation (microwave or resonator).

Regarding the rates for microwave transitions, there are two sharp thresholds given by the function $\Theta\bigl(\hbar \Omega - (\Delta \pm E_\mathrm{M})\bigr)$ for absorption and emission of photons at microwave frequencies $\Omega > 0$.
For instance, one QP from the continuum can decay to the MBS ($\Gamma_\mathrm{in,1}^\mathrm{mw}$) or can be promoted from it to the continuum ($\Gamma_\mathrm{out,1}^\mathrm{mw}$) for sufficient large energies $\hbar\Omega > \Delta - E_\mathrm{M}$.
For processes involving the ground state, we need to transfer two QPs. Either one continuum QP and the QP in the MBS combine to a Cooper pair by emission of a photon ($\Gamma_\mathrm{out,2}^\mathrm{mw}$) or a Cooper pair breaks up into two QPs, one is promoted to the MBS and one to the continuum, by photon absorption ($\Gamma_\mathrm{in,2}^\mathrm{mw}$). These transitions require energies $\hbar \Omega > \Delta + E_\mathrm{M}$.

The transitions involving photons exchanged with the damped LC resonator in the dc-SQUID can be discussed in a similar fashion, although there is no sharp threshold anymore due to a finite broadening of the resonator as shown in Eq.\,(\ref{eqn:spectralDensityResonator}). A finite environmental temperature $T_\mathrm{env} > 0$ allows the same four processes shown in Fig.\,\ref{fig:transitions} involving the resonator. 
For transitions involving single QPs, the amout of energy being absorbed (emitted) by the SSJ is $E - E_\mathrm{M}$ which is described by the rate $\Gamma_\mathrm{out,1}^\mathrm{res}$ ($\Gamma_\mathrm{in,1}^\mathrm{res}$). Moreover, the transition of two QPs, one from the MBS and one from the continuum, is described by the rates $\Gamma_\mathrm{out,2}^\mathrm{res}$ ($\Gamma_\mathrm{in,2}^\mathrm{res}$), in which an energy of $E + E_\mathrm{M}$ has to be emitted (absorbed). 

From Eq.\,(\ref{eqn:probabilitiesTJ}), we calculate the stationary occupation $n_\mathrm{M}$ of the MBS for $t \to \infty$. Using the property 
$P_0(t) + P_1(t) = 1$ together with $n_\mathrm{M}(t) = \mathrm{Tr} \{ \gamma_\mathrm{M}^{\dag} \gamma_\mathrm{M}^{\phantom{\dag}} \rho_\mathrm{M}(t) \}= P_1(t)$ for the topological junction, we find
\begin{equation}
n_\mathrm{M} = \frac{\Gamma_\mathrm{in}}{\Gamma_\mathrm{in} + \Gamma_\mathrm{out}} \, ,
\label{eqn:MBSoccupation}
\end{equation}
with the total in-/out-rate $\Gamma_\mathrm{in/out}$ as defined in Eq.\,(\ref{eqn:probabilitiesTJ}).

\subsection{Conventional junction}
\label{subsec:masterEqnCJ}
In contrast to the topological junction, the ABS in a conventional junction is twofold degenerate and can be occupied by two QPs. Therefore, the full set of equations for the probabilities $P_{0(2)}(t)$ of zero (double) occupancy and $P_1^{(j)}$ of single occupation reads
	\begin{equation}
		\frac{\d}{\d t} \begin{pmatrix} P_0(t) \\ P_1(t) \\ P_2(t) \end{pmatrix}
		=
		-\check{R}
		\begin{pmatrix} P_0(t) \\ P_1(t) \\ P_2(t) \end{pmatrix}
		\label{eqn:probabilitiesCJ}
	\end{equation}
with
\begin{equation}
	\check{R} = \begin{pmatrix} 
	 \Gamma_\mathrm{in,AA} + 2 \Gamma_\mathrm{in} & -\Gamma_\mathrm{out} &  -\Gamma_\mathrm{out,AA}
	\\ 
	-2 \Gamma_\mathrm{in} & \Gamma_\mathrm{in} + \Gamma_\mathrm{out} & -2 \Gamma_\mathrm{out}
	\\
	-\Gamma_\mathrm{in,AA} & -\Gamma_\mathrm{in} & \Gamma_\mathrm{out,AA} + 2 \Gamma_\mathrm{out}
	\end{pmatrix}\, .
	\label{eq:matrixRates}
\end{equation}
We have introduced the probability of single occupation as $P_1(t) = P_{1}^{(1)}(t) + P_{1}^{(2)}(t)$, as we cannot distinguish which zero-current state is occupied since the two states are symmetric\footnote{Except in the case discussed in Ref.\,[\onlinecite{Chtchelkatchev:2003ji}]}.
The rates given in the matrix in Eq.\,(\ref{eq:matrixRates}) are defined as
\begin{subequations}
	\begin{align}
	\Gamma_\mathrm{in} &= \Gamma_\mathrm{in,1}^\mathrm{mw} + \Gamma_\mathrm{in,1}^\mathrm{res} + \Gamma_\mathrm{in,2}^\mathrm{mw} + \Gamma_\mathrm{in,2}^\mathrm{res} \, , \\
	\Gamma_\mathrm{out} &= \Gamma_\mathrm{out,1}^\mathrm{mw} + \Gamma_\mathrm{out,1}^\mathrm{res} + \Gamma_\mathrm{out,2}^\mathrm{mw} + \Gamma_\mathrm{out,2}^\mathrm{res} \, , \\
	\Gamma_\mathrm{in,AA} &= \Gamma_\mathrm{in,AA}^\mathrm{mw} + \Gamma_\mathrm{in,AA}^\mathrm{res} \, , \\
	\Gamma_\mathrm{out,AA} &= \Gamma_\mathrm{out,AA}^\mathrm{mw} + \Gamma_\mathrm{out,AA}^\mathrm{res} \, .
	\end{align}
	\label{eqn:InOutDoubleInOut}%
\end{subequations}
The transition matrix elements obtained from the current operator in Eq.\,(\ref{eqn:currentOperatorSSJ}) in the case of a conventional SSJ are explicitly shown in appendix \ref{subappendix:b2}. As before, we define an effective density of states
\begin{multline}
\rho_\cj^\pm(E) = \frac{\sqrt{\Delta^2 - E_\mathrm{A}^2}}{\Delta^2} \, \frac{\sqrt{E^2 - \Delta^2}}{E^2 - E_\mathrm{A}^2} \\ 
\times \frac{E_\mathrm{A} (E \pm E_\mathrm{A}) \mp \Delta^2 (\cos\varphi + 1)}{E_\mathrm{A}}  \qquad 
\end{multline}
for the conventional junction. 
The individual rates entering Eq.\,(\ref{eqn:InOutDoubleInOut}) due to the microwave read
\begin{subequations}
	\begin{align}
	\Gamma_\mathrm{out,2/in,1}^\mathrm{mw} &= \frac{(\delta\varphi)^2 \Delta^2}{32 \hbar} \, \mathcal T \, \rho_\cj^\pm(\hbar \Omega \mp E_\mathrm{A})
	\nonumber  \\
	& \qquad  \times  f(\hbar \Omega \mp E_\mathrm{A}) \, \Theta\bigl(\hbar \Omega - (\Delta \pm E_\mathrm{A})\bigr) , \\
	\Gamma_\mathrm{in,2/out,1}^\mathrm{mw} &= \frac{(\delta\varphi)^2 \Delta^2}{32 \hbar} \, \mathcal T \, \rho_\cj^\pm(\hbar \Omega \mp E_\mathrm{A}) \,  \nonumber \\
	&  \times \bigl( 1-f(\hbar \Omega \mp E_\mathrm{A}) \bigr) \, \Theta\bigl(\hbar  \Omega - (\Delta \pm E_\mathrm{A}) \bigr) , \label{eqn:microwaveAbsorptionCJ}
	\end{align}
	\label{eqn:microwaveConventional}%
\end{subequations}
whereas we have
\begin{subequations}
	\begin{align}
	\Gamma_\mathrm{out,2/in,1}^\mathrm{res} &= \frac{\lambda^2 \Delta^2}{32 \hbar} \, \mathcal T  \int_{\Delta}^{\infty} \!\!\!\!  \d E \, \rho_\cj^\pm(E) \,  f(E) 
	\nonumber \\
	& \qquad  \times   \chi(E \pm E_\mathrm{A}) \, \bigl( 1+n_\mathrm{B}(E \pm E_\mathrm{A}) \bigr) , \label{eqn:resonatorRatesZeroTemp}\\
	\Gamma_\mathrm{in,2/out,1}^\mathrm{res} &= \frac{\lambda^2 \Delta^2}{32 \hbar} \, \mathcal T  \int_{\Delta}^{\infty} \!\!\!\! \d E \, \rho_\cj^\pm(E) \, \bigl( 1-f(E) \bigr) 
	\nonumber \\
	& \qquad \quad \quad  \times  \chi(E \pm E_\mathrm{A}) \, n_\mathrm{B}(E \pm E_\mathrm{A}) \, ,
	\end{align}
	\label{eq:resonatorAndreev}%
\end{subequations}
for the photons exchanged with the resonator. 
For the case of the conventional junction, there are additional parity-conserving transitions proportional to the reflection $(1-\mathcal T)$ at the interface which describe the excitation (relaxation) of two QPs with $2E_\mathrm{A}$ from (to) the ground state
\begin{subequations}
	\begin{align}
	\Gamma_{b\mathrm{,AA}}^\mathrm{mw} &= \frac{(\delta\varphi)^2 \Delta^3}{32 \hbar} \, (1-\mathcal T) \, \rho_\cj^\mathrm{A}(E_\mathrm{A}) \,  S_\mathrm{ph}(\hbar\Omega - 2E_\mathrm{A}) , \label{eqn:AndreevDoubleRateMicrowave} \\
	\Gamma_{b\mathrm{,AA}}^\mathrm{res} &=  \frac{\lambda^2 \Delta^3}{32 \hbar} \, (1-\mathcal T) \, \rho_\cj^\mathrm{A}(E_\mathrm{A}) \, \chi(2E_\mathrm{A}) \,  \bigl(\delta_{b,\mathrm{out}} + n_\mathrm{B}(2E_\mathrm{A}) \bigr) , \label{eqn:AndreevDoubleRate}
	\end{align}
	\label{eqn:doubleRates}%
\end{subequations}
with $b \in \{ \mathrm{in,out} \}$ and an effective density of states for the bound states
\begin{equation}
\rho_\cj^\mathrm{A}(E_\mathrm{A}) = \frac{\pi}{\Delta^3} \frac{(\Delta^2 - E_\mathrm{A}^2)^2}{E_\mathrm{A}^2} \, . \qquad 
\end{equation}
These rates are associated with a transfer of a Cooper pair between the twofold degenerate Andreev level and the ground state. In passing by, we have introduced a phenomenological Lorentzian broadening $S_\mathrm{ph}(E) = (\gamma_\mathrm{A}/\pi)/(E^2 + \gamma_\mathrm{A}^2)$ of the Andreev level with width $\gamma_\mathrm{A}$ in the rates $\Gamma_\mathrm{in/out,AA}^\mathrm{mw}$ to resolve the transition of a Cooper pair between the ground state and the ABS\cite{Kos:2013hl}.

We notice that the rates changing the parity for the conventional junction differ from the ones of the topological junction by the factor transmission $\mathcal T$, beyond the obvious substitution $\rho_\tj \to \rho_\cj$.
For $\mathcal T = 1$, the two junctions are exactly equivalent and the occupations are related by $n_\mathrm{A} = 2 n_\mathrm{M}$ where the factor of 2 is due to the remaining spin-degeneracy.
For $T_\mathrm{qp} = 0$, our rates in Eq.\,(\ref{eqn:microwaveAbsorptionCJ}) 
coincide with the ones calculated in Ref.\,[\onlinecite{Riwar:2014ih}] and with 
the calculation of the admittance $Y(\Omega)$ in a short conventional junction in Ref. [\onlinecite{Kos:2013hl}].
Moreover, at $T_\mathrm{env} = 0$, the parity-conserving rate $\Gamma_{\mathrm{out,AA}}^\mathrm{res}$ in Eq.\,(\ref{eqn:AndreevDoubleRate}) has the same form as the annihilition rate found in Ref.\,[\onlinecite{Riwar:2014ih}]. All rates due to the resonator (Eqs.\,\,(\ref{eq:resonatorAndreev}) and (\ref{eqn:AndreevDoubleRate})) coincide with the rates found in Ref.\,[\onlinecite{Zazunov:2014kn}].

The discussion of the in- and out-rates involving continuum QPs is analog to the case of a topological junction.
In addition, there are new processes which directly switch the occupation of the ABS without changing the parity 
(cf.\,Eq.\,(\ref{eqn:doubleRates})). 
These rates appear in any short conventional junction as long as the transmission is $\mathcal T < 1$. Since no QPs from the continuum are involved in these rates, they are completely independent of $T_\mathrm{qp}$ and they occur at an energy of $2E_\mathrm{A}$.
In the case of microwave emission (absorption) $\Gamma_{\mathrm{out(in),AA}}^\mathrm{mw}$, the microwave energy 
has to be $\hbar \Omega \approx 2E_\mathrm{A}$ for the process to be in resonance. In the same way the resonator has to provide energies of $2 E_\mathrm{A}$ in order to promote a Cooper pair to the ABS according to the rate $\Gamma_{\mathrm{in,AA}}^\mathrm{res}$. 

From Eq.\,(\ref{eqn:probabilitiesCJ}), we calculate the stationary occupation $n_\mathrm{A}$ of the ABS for $t \to \infty$. Using the property $P_0(t) + P_1(t) + P_2(t) = 1$ together with $n_\mathrm{A}(t) = \mathrm{Tr} \{ (\gamma_\mathrm{A,1}^{\dag} \gamma_\mathrm{A,1}^{\phantom{\dag}} + \gamma_\mathrm{A,2}^{\dag} \gamma_\mathrm{A,2}^{\phantom{\dag}}) \rho_\mathrm{A}(t) \}$ for the conventional junction, we eliminate the probability $P_1(t)$ and find
\begin{equation}
	n_\mathrm{A} = 1 + P_2 - P_0 \, ,
	\label{eqn:occupationABS}
\end{equation}
with $P_2$ ($P_0$) being the stationary probability for the bound state of being occupied with two (zero) QPs. For these probabilities, we find the relations
\begin{equation}
\begin{pmatrix} P_0 \\ P_2 \end{pmatrix} 
= \frac{\check{\Gamma}}{\mathrm{det}\,\check\Gamma}   \begin{pmatrix} \Gamma_\mathrm{out} \\ \Gamma_\mathrm{in} \end{pmatrix} \, ,
\end{equation}
with the matrix 
\begin{equation}
	\check{\Gamma} 
	= 
	\begin{pmatrix} 
	\Gamma_\mathrm{out,AA} + \Gamma_\mathrm{in} + 2 \,  \Gamma_\mathrm{out} & \Gamma_\mathrm{out,AA} - \Gamma_\mathrm{out} \\
	\Gamma_\mathrm{in,AA} - \Gamma_\mathrm{in} & \Gamma_\mathrm{in,AA}  + \Gamma_\mathrm{out} + 2 \, \Gamma_\mathrm{in} \\
	\end{pmatrix}
	\label{eqn:RateMatrix}
\end{equation}
and the corresponding rates as defined in Eq.\,(\ref{eqn:InOutDoubleInOut}).

%
%
%
\section{Results}
\label{sec:results}

\begin{figure*}
	\centering
	\includegraphics[width=0.9\linewidth]{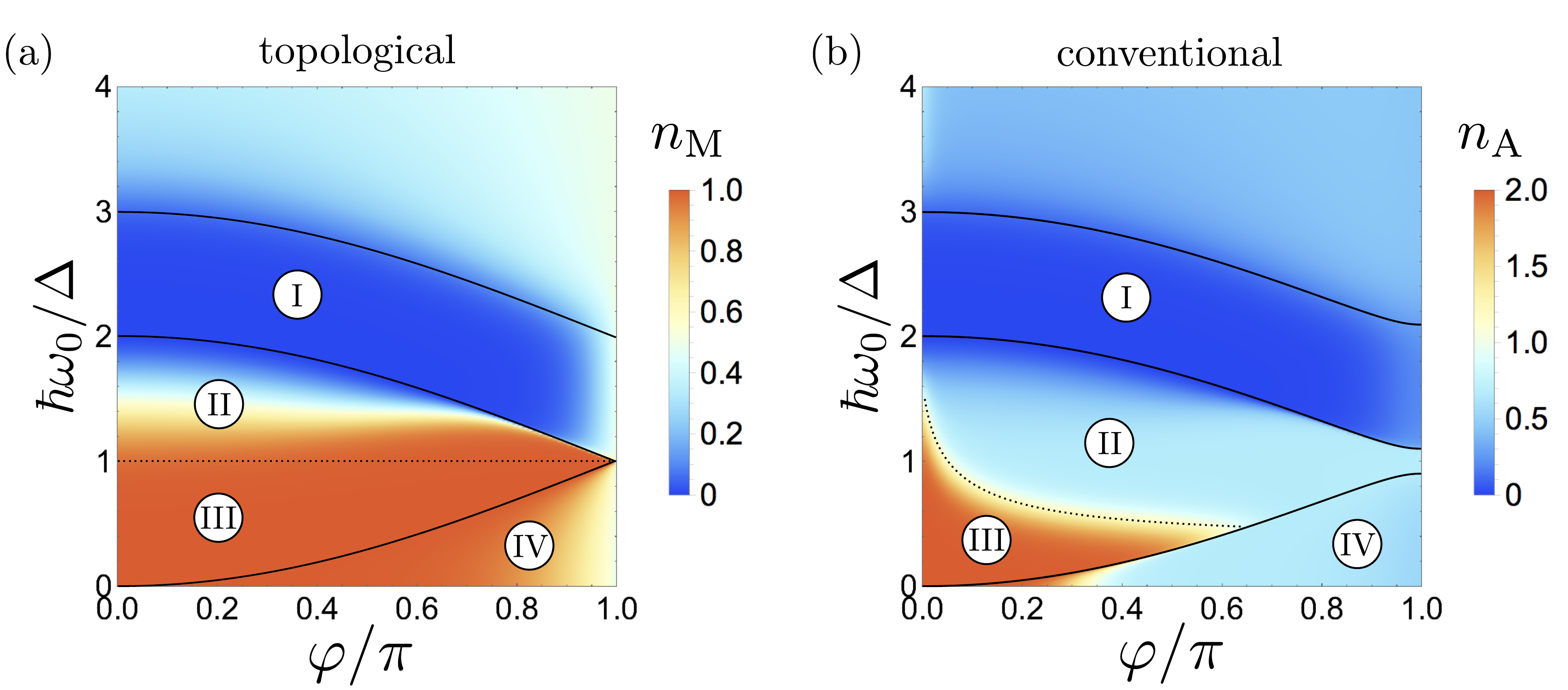}
	\caption{Stationary nonequilibrium bound state occupation in SSJ as a function of phase difference $\varphi$ and Josephson plasma frequency $\omega_0$. 
	(a) MBS occupation $n_\mathrm{M}$ in topological junctions. The black solid lines separating regions (I) and (II) is given by $\hbar \omega_0 = \Delta + E_\mathrm{M}$ 
and by $\hbar \omega_0 = \Delta - E_\mathrm{M}$ for the region (III) and (IV). 
The black dashed line separating regions (II) and (III) is $\hbar \omega_0 = \Delta$. The upper bound for region (I) is given by $\hbar \omega_0 = 2 \Delta + E_\mathrm{M}$.
(b) ABS occupation $n_\mathrm{A}$ for finite transmission $\mathcal T = 0.99$ in conventional junctions. 
The black solid line separating regions (I) and (II)  is given by $\hbar \omega_0 = \Delta + E_\mathrm{A}$ 
and by $\hbar \omega_0 = \Delta - E_\mathrm{A}$ for the region (II)/ (III) and (IV).
The black dashed line separating regions (II) and (III) is given by $\Gamma_{\mathrm{out,AA}}^\mathrm{res} \approx 2 \Gamma_\mathrm{in,1}^\mathrm{res}$ in the limit $\Gamma_{\mathrm{out,AA}}^\mathrm{res} , \Gamma_\mathrm{in,1}^\mathrm{res} \gg \Gamma_\mathrm{out,2}^\mathrm{res}$. The upper bound for region (I) is given by $\hbar \omega_0 = 2 \Delta + E_\mathrm{A}$. Common parameters: $\gamma = 0.001 \Delta$, $\kb T_\mathrm{qp} = 0.1 \Delta$, $T_\mathrm{env} = 0$.}
	\label{fig:occupationResonator}
\end{figure*}
We study the case in which we have two different temperatures in the system. The environmental temperature $T_\mathrm{env}$, which is considered to be the temperature at which the experiment is performed, i.e. the resonator, and a QP temperature $T_\mathrm{qp}$. This is motivated by experiments on superconducting low temperature circuits where the continuum QP population does not correspond to thermal equilibrium.

\subsection{Effect of the damped LC-resonator}
\label{subsec:noMicrowave}
First, we investigate the effect of the damped resonator on the occupation of the bound state in absence of any microwave radiation. 
As a first observation, we note that if both temperatures are equal, i.e. $T_\mathrm{env} = T_\mathrm{qp} = T$, one has $\Gamma_\mathrm{out}^\mathrm{res} / \Gamma_\mathrm{in}^\mathrm{res} = e^{E_\alpha / T}$ (detailed balance) for both the topological ($\alpha = \mathrm{M}$) and the conventional ($\alpha = \mathrm{A}$) junction. In this case, the stationary solutions of the bound state occupations reduce to $n_\mathrm{M} = f(E_\mathrm{M})$ and $n_\mathrm{A} = 2 f(E_\mathrm{A})$ for the topological and the conventional junction, respectively, with the Fermi function $f(E)$ at the equilibrium temperature $T$.

From now on, we consider different temperatures for continuum QPs and the environment, $T_\mathrm{env} \neq T_\mathrm{qp}$. 

We plot the occupation as a function of the phase bias $\varphi$ and the Josephson plasma frequency $\omega_0$ at zero environmental temperature $T_\mathrm{env} = 0$,
(namely $\kb T_\mathrm{env}  \ll \hbar \omega_0$). 
In this limit, $n_\mathrm{B}(E) = 0$ and the resonator is only able to absorb energy emitted by transitions of QPs in the SSJ. 
The expressions for the occupation of the MBS and the ABS, Eqs.\,(\ref{eqn:MBSoccupation}) and (\ref{eqn:occupationABS}), respectively, reduce to
\begin{align}
	n_\mathrm{M} = \frac{\Gamma_\mathrm{in,1}^\mathrm{res}}{\Gamma_\mathrm{in,1}^\mathrm{res} + \Gamma_\mathrm{out,2}^\mathrm{res}} 
	\label{eqn:MBSreduced}
\end{align}
and
\begin{equation}
	n_\mathrm{A} = \frac{2 \Gamma_\mathrm{in,1}^\mathrm{res} \bigl( 2 (\Gamma_\mathrm{in,1}^\mathrm{res} + \Gamma_\mathrm{out,2}^\mathrm{res}) + \Gamma_{\mathrm{out,AA}}^\mathrm{res} \bigr)}{\Gamma_{\mathrm{out,AA}}^\mathrm{res} (3 \Gamma_\mathrm{in,1}^\mathrm{res} + \Gamma_\mathrm{out,2}^\mathrm{res}) + 2 (\Gamma_\mathrm{in,1}^\mathrm{res} + \Gamma_\mathrm{out,2}^\mathrm{res})^2} \, ,
\end{equation}
respectively.
As expected in the limit $\Gamma_{\mathrm{out,AA}}^\mathrm{res} \ll \Gamma_\mathrm{in,1}^\mathrm{res} , \Gamma_\mathrm{out,2}^\mathrm{res}$, the two junctions behave in a similar way and the occupation is simply rescaled, $n_\mathrm{A} \approx 2 n_\mathrm{M}$. Hence, the different behaviour relies on the presence of the process $\Gamma_{\mathrm{out,AA}}^\mathrm{res}$ (red dotted arrows in Fig.\,\ref{fig:transitions}). The results are shown in Fig.\,\ref{fig:occupationResonator} at $\kb T_\mathrm{qp} = 0.1 \Delta$.

\begin{figure}
	\centering
	\includegraphics[width=1\linewidth]{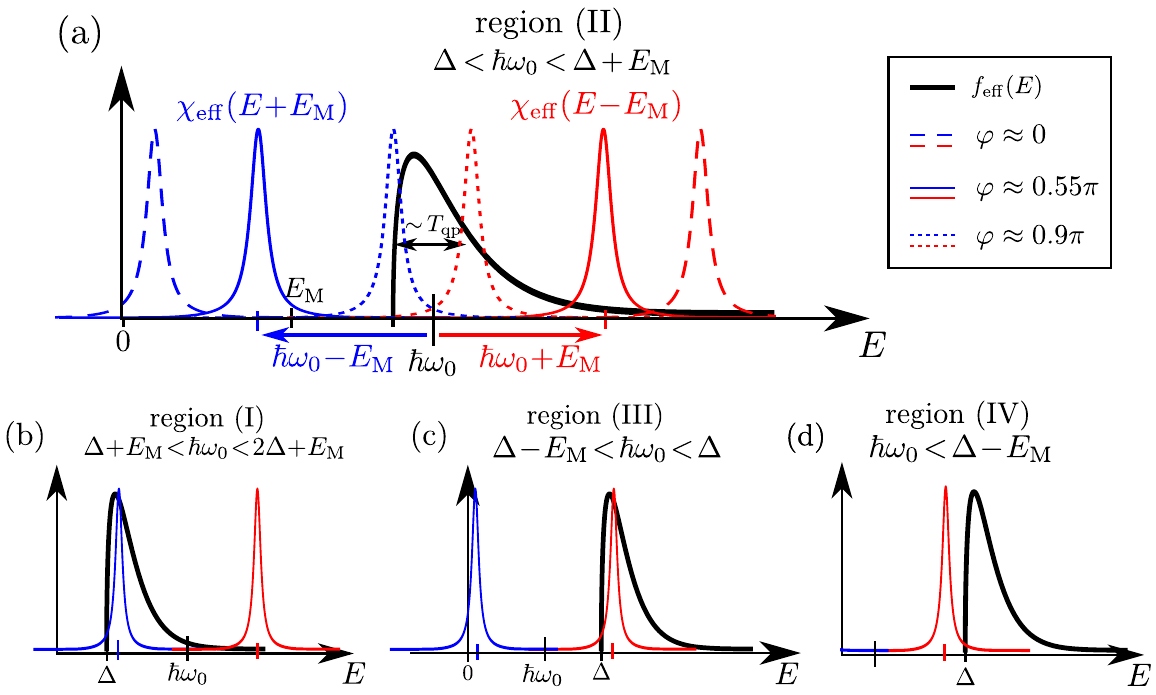}
	\caption{Sketch of the contributions to the transition rates $\Gamma_\mathrm{in,1/out,2}^\mathrm{res}$ at $T_\mathrm{env} = 0$ which determine the absolute value of the rates depending on the value of the Josephson plasma frequency $\omega_0$. QPs in the continuum with energies $E > \Delta$ are occupied by the effective function $f_\mathrm{eff}(E) = f(E) \sqrt{E^2-\Delta^2}/\Delta$, which is a combination of matrix elements, density of states in a superconductor and the Fermi-Dirac distribution. Most of the continuum QPs are located closely above the gap in an energy range approximately proportional to $T_\mathrm{qp}$. The effective spectral function $\chi_\mathrm{eff}(E) = \chi(E) \Delta/E$ is related to the absorption peak of the resonator shifted to values $\omega_0 \pm E_\mathrm{M}$ according to $\chi_\mathrm{eff}(E \mp E_\mathrm{M})$ for the rate $\Gamma_\mathrm{in,1/out,2}^\mathrm{res}$. The maximal shifts $\omega_0 \pm \Delta$ can be achieved at a phase difference $\varphi = 0$, while the minimal shifts are at $\varphi = \pi$. The cases (a), (b), (c) and (d) correspond to the regions (II), (I), (III) and (IV), respectively, defined in Fig.\,\ref{fig:occupationResonator}(a).}
	\label{fig:sketchOfParityChangingRates}
\end{figure}

Next, we discuss the occupation in case of a topological SSJ shown in Fig.\,\ref{fig:occupationResonator}(a). 
The rates entering $n_\mathrm{M}$ in Eq.\,(\ref{eqn:MBSreduced}) are given by
\begin{align}
	\Gamma_\mathrm{out,2/in,1}^\mathrm{res} \sim \int_{\Delta}^{\infty} \!\!\! \d E \, f_\mathrm{eff}(E) \, \chi_\mathrm{eff}(E \pm E_\mathrm{M}) \, ,
\end{align}
where we defined the effective occupation function $f_\mathrm{eff}(E) = f(E) \sqrt{E^2 - \Delta^2} / \Delta$ and the effective spectral density $\chi_\mathrm{eff}(E) = \chi(E) \Delta/E$.
For $\hbar \omega_0 > \Delta$, we distinguish between two regions: region (I), with $\Delta + E_\mathrm{M} < \hbar \omega_0 < 2\Delta + E_\mathrm{M}$, showing an empty MBS, and region (II), with $\Delta < \hbar \omega_0 < \Delta + E_\mathrm{M}$, in which the occupation of the MBS varies strongly.
This can be understood by means of Fig.\,\ref{fig:sketchOfParityChangingRates}. The absolute value of the rates is determined by the convolution of the effective functions $f_\mathrm{eff}(E)$ and $\chi_\mathrm{eff}(E \pm E_\mathrm{M})$. 

For region (I), there is a strong overlap between $f_\mathrm{eff}(E)$ and $\chi_\mathrm{eff}(E + E_\mathrm{M})$ in the emission rate $\Gamma_\mathrm{out,2}^\mathrm{res}$ due to the location of the majority of the continuum QPs closely above the gap leading to $\Gamma_\mathrm{out,2}^\mathrm{res} \gg \Gamma_\mathrm{in,1}^\mathrm{res}$ (cf. Fig.\,\ref{fig:sketchOfParityChangingRates}(b)). Therefore, the interaction with the resonator leads to a strong depopulation of the MBS with $n_\mathrm{M} \ll 1$.
In region (II), the occupation strongly depends on the value of $\omega_0$ and the phase $\varphi$. By looking at Fig.\,\ref{fig:sketchOfParityChangingRates}(a), we start to decrease the value of $\omega_0$ from $\hbar\omega_0 = \Delta + E_\mathrm{M}$ to $\hbar \omega_0 = \Delta$. A value of $\hbar \omega_0 \lesssim \Delta + E_\mathrm{M}$ shows that there is a large overlap for $f_\mathrm{eff}(E)$ and $\chi_\mathrm{eff}(E + E_\mathrm{M})$, while the overlap between $f_\mathrm{eff}(E)$ and $\chi_\mathrm{eff}(E - E_\mathrm{M})$ is negligible. By decreasing $\omega_0$, we shift the peaks of $\chi_\mathrm{eff}(E \pm E_\mathrm{M})$ to lower energies and, in particular, $\chi_\mathrm{eff}(E + E_\mathrm{M})$ below the gap $\Delta$ which drastically reduces the previously large overlap, eventually leading to $\Gamma_\mathrm{in,1}^\mathrm{res} \gg \Gamma_\mathrm{out,2}^\mathrm{res}$ and a highly populated MBS.

Decreasing the Josephson plasma frequency further, i.e. $\hbar \omega_0 < \Delta$ (regions (III) and (IV)), the depopulating rate $\Gamma_\mathrm{out,2}^\mathrm{res}$ becomes negligible everywhere except for phase differences $\varphi \approx \pi$ in which $\Gamma_\mathrm{out,2}^\mathrm{res} \approx \Gamma_\mathrm{in,1}^\mathrm{res}$ (region (IV)). Therefore, coupling to the resonator leads to a high occupation $n_\mathrm{M} \approx 1$ of the MBS.

In the case of the conventional junction, the discussion of the occupation of the ABS shown in Fig.\,\ref{fig:occupationResonator}(b) is analogous to the case of the topological junction, albeit the behaviour is more complex due to the presence of the process $\Gamma_\mathrm{out,AA}^\mathrm{res}$.

For region (I), with $\Delta + E_\mathrm{A} < \hbar \omega_0 < 2\Delta + E_\mathrm{A}$, the populating rate $\Gamma_\mathrm{in,1}^\mathrm{res}$ is again stronlgy suppressed and the occupation of the ABS becomes $n_\mathrm{A} \ll 1$, similar to the case of the topological junction.
The other regions (II), (III) and (IV) can be explained by the competition between the relaxation process associated with the rate $\Gamma_\mathrm{out,AA}^\mathrm{res}$ and the refilling process involving continuum QPs with rate $\Gamma_\mathrm{in,1}^\mathrm{res}$.
For $\Delta < \hbar \omega_0  < \Delta + E_\mathrm{A}$, which is mainly region (II), transitions associated with the rate $\Gamma_\mathrm{out,AA}^\mathrm{res}$ dominate the behaviour of the occupation, i.e. $\Gamma_\mathrm{out,AA}^\mathrm{res} \gg \Gamma_\mathrm{in,1}^\mathrm{res},\Gamma_\mathrm{out,2}^\mathrm{res}$. Therefore, the occupation reduces to
\begin{equation}
	n_\mathrm{A} \approx \frac{2 \Gamma_\mathrm{in,1}^\mathrm{res}
	}{
		3 \Gamma_\mathrm{in,1}^\mathrm{res} + \Gamma_\mathrm{out,2}^\mathrm{res}} \, .
\end{equation}
Due to the presence of the process $\Gamma_\mathrm{out,AA}^\mathrm{res}$, there is a strong reduction of the occupation of the ABS compared to the topological case leading to an occupation of $n_\mathrm{A} \approx 2/3$ for $\Gamma_\mathrm{in,1}^\mathrm{res} \gg \Gamma_\mathrm{out,2}^\mathrm{res}$.

At energies $\hbar \omega_0  < \Delta$, mainly region (III), (IV) and partially region (II), transitions due to the process $\Gamma_\mathrm{out,2}^\mathrm{res}$ are negligible and, therefore, the two competing rates are $\Gamma_\mathrm{out,AA}^\mathrm{res}$ and $\Gamma_\mathrm{in,1}^\mathrm{res}$. The occupation reduces to
\begin{equation}
	n_\mathrm{A} \approx \frac{
		4 \Gamma_\mathrm{in,1}^\mathrm{res} + 2 \Gamma_\mathrm{out,AA}^\mathrm{res}
	}{
		2 \Gamma_\mathrm{in,1}^\mathrm{res} + 3 \Gamma_\mathrm{out,AA}^\mathrm{res}
	} \, .
\end{equation}
The regions (II) and (III), for which $\Delta - E_\mathrm{A} < \hbar \omega_0 < \Delta$, are separated by the dashed line $n_\mathrm{A} \approx 1$ for $\Gamma_\mathrm{out,AA}^\mathrm{res} \approx 2 \Gamma_\mathrm{in,1}^\mathrm{res}$ with either $n_\mathrm{A} \lesssim 2$ for $\Gamma_\mathrm{out,AA}^\mathrm{res} \ll \Gamma_\mathrm{in,1}^\mathrm{res}$ (region (III)) or $n_\mathrm{A} \approx 2/3$ for $\Gamma_\mathrm{out,AA}^\mathrm{res} \gg \Gamma_\mathrm{in,1}^\mathrm{res}$ (region (II)).
For $\hbar \omega_0 < \Delta - E_\mathrm{A}$, we find again $\Gamma_\mathrm{out,AA}^\mathrm{res} \gg \Gamma_\mathrm{in,1}^\mathrm{res}$ leading to an occupation of $n_\mathrm{A} \approx 2/3$. This is a consequence of the lack of continuum QPs at energies which can be absorbed by the resonator in order to refill the bound state.
The reason that $\Gamma_\mathrm{out,AA}^\mathrm{res}  \gg \Gamma_\mathrm{in,1}^\mathrm{res}, \Gamma_\mathrm{out,2}^\mathrm{res}$ almost everywhere except the red region is due to the fact that the absolute value of $\Gamma_\mathrm{out,AA}^\mathrm{res}$ is independent of the continuum. In contrast, $\Gamma_\mathrm{in,1}^\mathrm{res}$ and $\Gamma_\mathrm{out,2}^\mathrm{res}$ consist of a convolution of functions of the continuum and the absorption of the resonator.

\begin{figure}
	\includegraphics[width=0.85\linewidth]{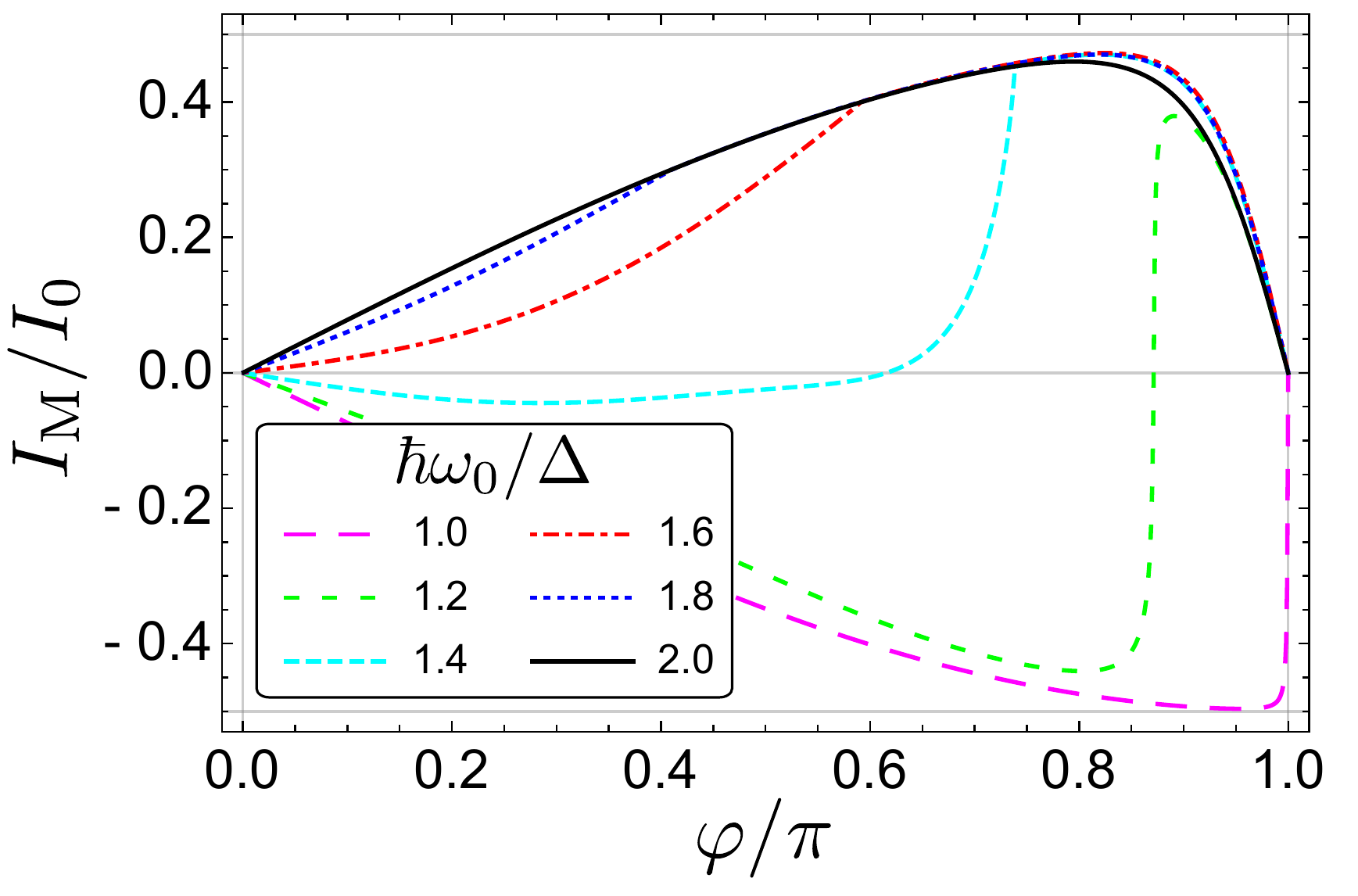}
	\caption{Supercurrent $I_\mathrm{M}$ carried by Majorana bound states in short topological superconducting junctions as a function of phase difference $\varphi$ for different Josephson plasma frequencies $\omega_0$. Parameters are the same as for Fig.\,\ref{fig:occupationResonator}(a). Current is plotted in units of $I_0 = e \Delta / \hbar$.}
	\label{fig:CurrentPhaseRelationMBS1}
\end{figure}

\begin{figure}
	\centering
	\includegraphics[width=0.87\linewidth]{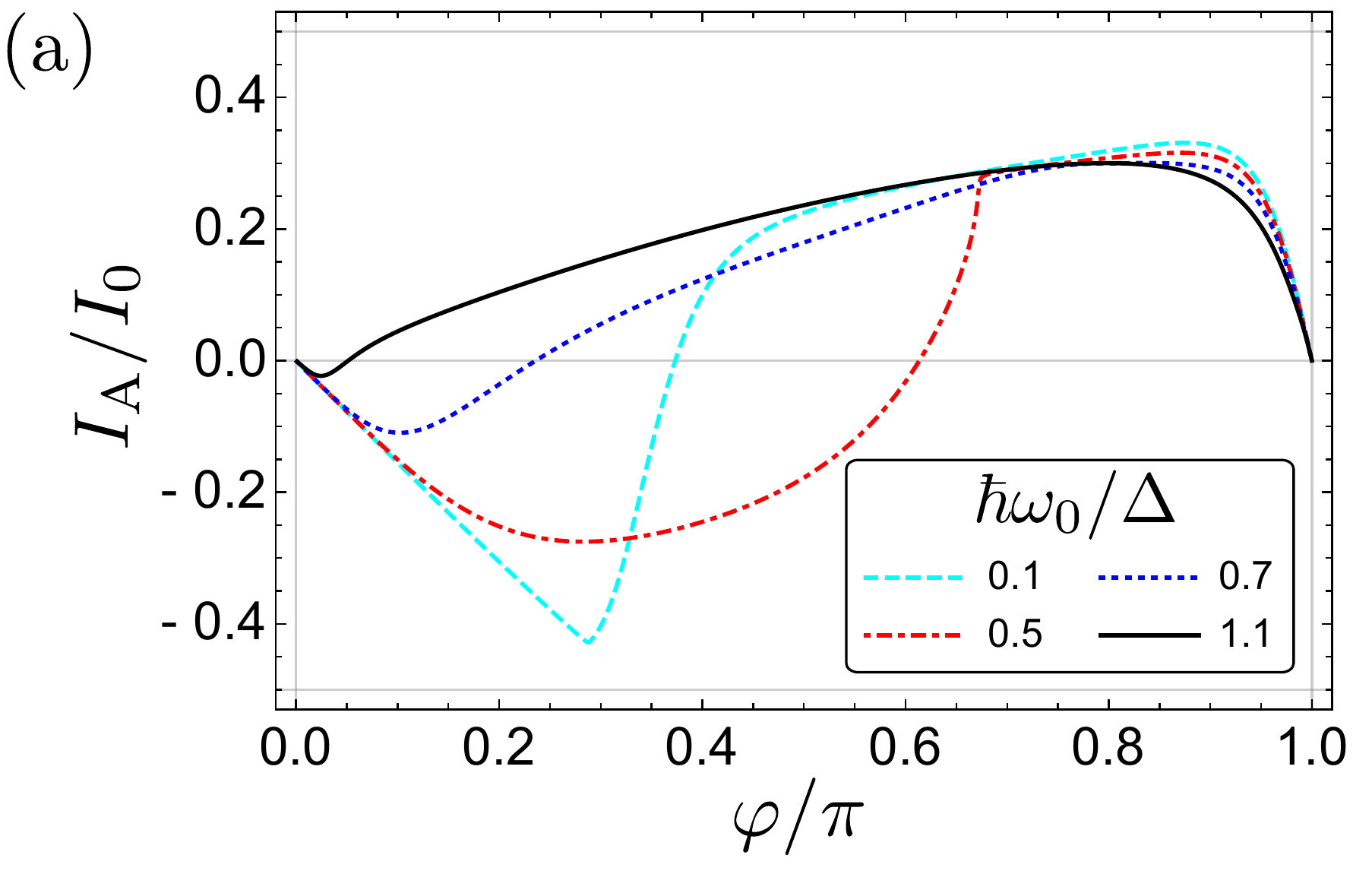}
	\includegraphics[width=0.87\linewidth]{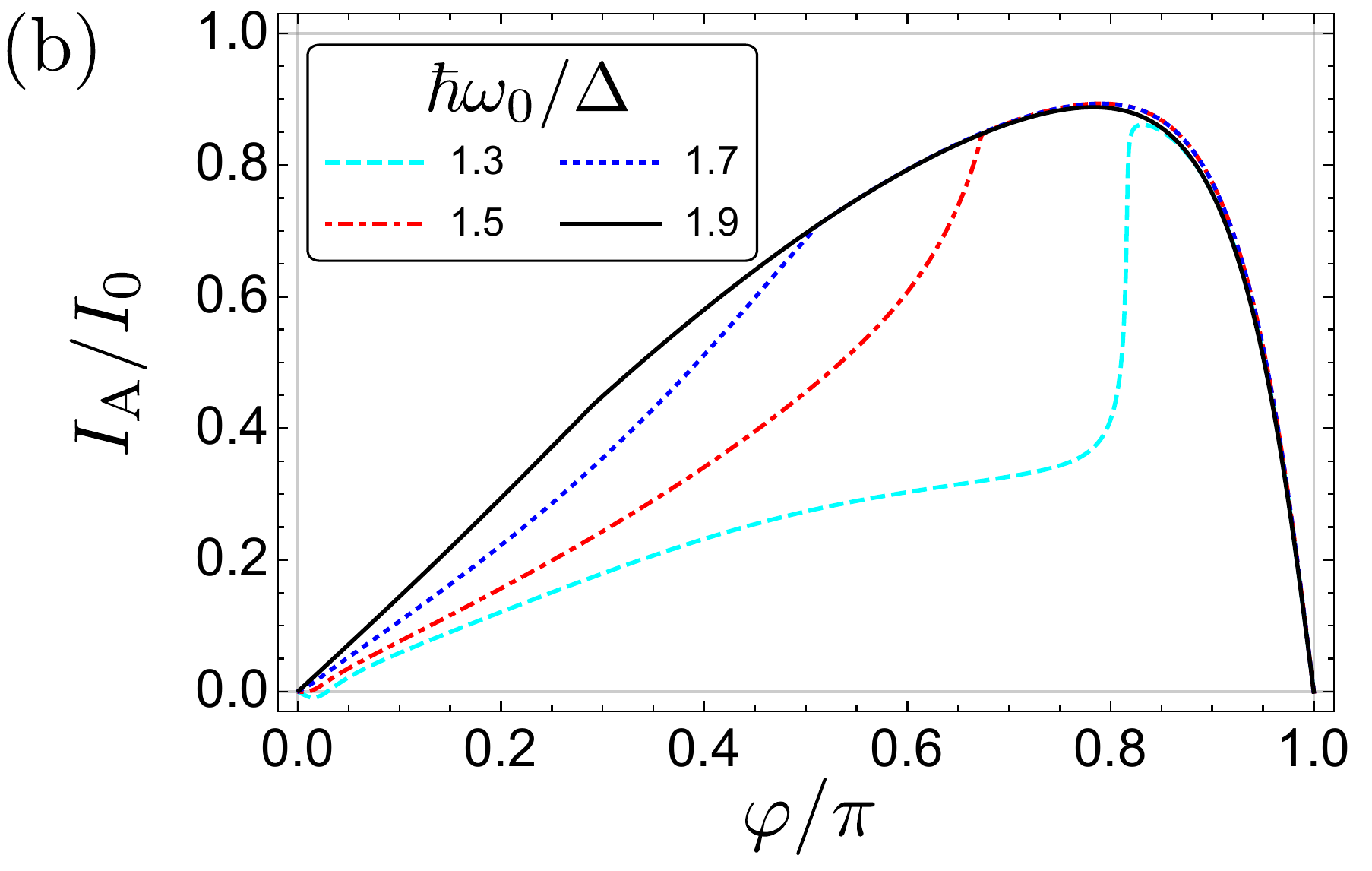}
	\caption{Supercurrent $I_\mathrm{A}$ carried by Andreev bound states in short conventional superconducting junctions as a function of phase difference $\varphi$ for Josephson plasma frequencies: (a) $\hbar \omega_0 \leq 1.1\Delta$ and (b) $\hbar \omega_0 \geq 1.3\Delta$. Parameters are the same as for Fig.\,\ref{fig:occupationResonator}(b). Current is plotted in units of $I_0 = e \Delta / \hbar$.}
	\label{fig:CurrentPhaseRelationABS1}
\end{figure}

As described in Sec.\,\ref{subsec:ssj}, both MBS and ABS carry 
a supercurrent $I_\mathrm{M}$ and $I_\mathrm{A}$, respectively, which depends on the occupation of the bound state (cf. Eqs.\,(\ref{eqn:MBScurrent}) and (\ref{eqn:ABScurrent}), respectively).
In the case of a topological junction, the occupation $n_\mathrm{M}$ shows a nontrivial behaviour for $\Delta \leq \hbar \omega_0 \leq 2 \Delta$ (cf. Fig.\,\ref{fig:occupationResonator}(a)). The corresponding current $I_\mathrm{M}$ as a function of the phase difference $\varphi$ for fixed $\omega_0$ in this range is shown in Fig.\,\ref{fig:CurrentPhaseRelationMBS1}. 
For $\varphi = 0$ and $\varphi = \pi$, the corresponding current is always zero which is independent of the value of $\omega_0$ since either $\partial E_\mathrm{M}/\partial \varphi |_{\varphi = 0} = 0$ or 
$n_\mathrm{M}(\varphi = \pi) = 1/2$. 
Moreover, for $\varphi \neq 0, \pi$, there exists another zero current crossing for $\Delta < \hbar \omega_0 \lesssim 1.5 \Delta$ since the occupation crosses $n_\mathrm{M} = 1/2$ leading to a change of the direction of the current across this junction. 
For $\hbar \omega_0 \gtrsim 1.5 \Delta$ and $\hbar \omega_0 \leq \Delta$, the occupation is always $n_\mathrm{M} < 1/2$ and $n_\mathrm{M}> 1/2$, respectively, such that the additional zero current crossing disappears.

In the case of a conventional junction, the occupation $n_\mathrm{A}$ shows a nontrivial behaviour for $\hbar \omega_0 < 2 \Delta$. 
In this regime, we plot the corresponding current $I_\mathrm{A}$ as a function of the phase difference $\varphi$ for fixed $\omega_0$, shown in Fig.\,\ref{fig:CurrentPhaseRelationABS1}(a) for $\hbar \omega_0 \leq 1.1\Delta$ and in Fig.\,\ref{fig:CurrentPhaseRelationABS1}(b) for $\hbar \omega_0 \geq 1.3\Delta$.
For $\varphi = 0$ and $\varphi = \pi$, the corresponding current is always zero which is independent of the value of $\omega_0$, similar to the topological case, yet both zero crossings originate from $\partial E_\mathrm{A}/\partial \varphi |_{\varphi = 0,\pi} = 0$. 
For phase differences $\varphi \neq 0,\pi$, there exists an additional zero crossing if the occupation of the ABS crosses $n_\mathrm{A} = 1$.
We observe such zero current crossing for $\hbar \omega_0 \leq \Delta$ due to the presence of the rate $\Gamma_\mathrm{out,AA}^\mathrm{res}$ in a highly transmitting contact, at finite value of the phase.
Eventually, the crossing point of zero current shifts to very  small phase differences 
$\varphi \ll \pi$ at $\Delta < \hbar \omega_0 \lesssim 1.5\Delta$. 
For values $\hbar \omega_0 \gtrsim 1.5\Delta$, this additional zero current crossing disappears 
since the occupation is  always $n_\mathrm{A} < 1$.

\subsection{LC resonator and microwave}
\label{subsec:withMicrowave}

\begin{figure}
	\centering
	\includegraphics[width=0.99\linewidth]{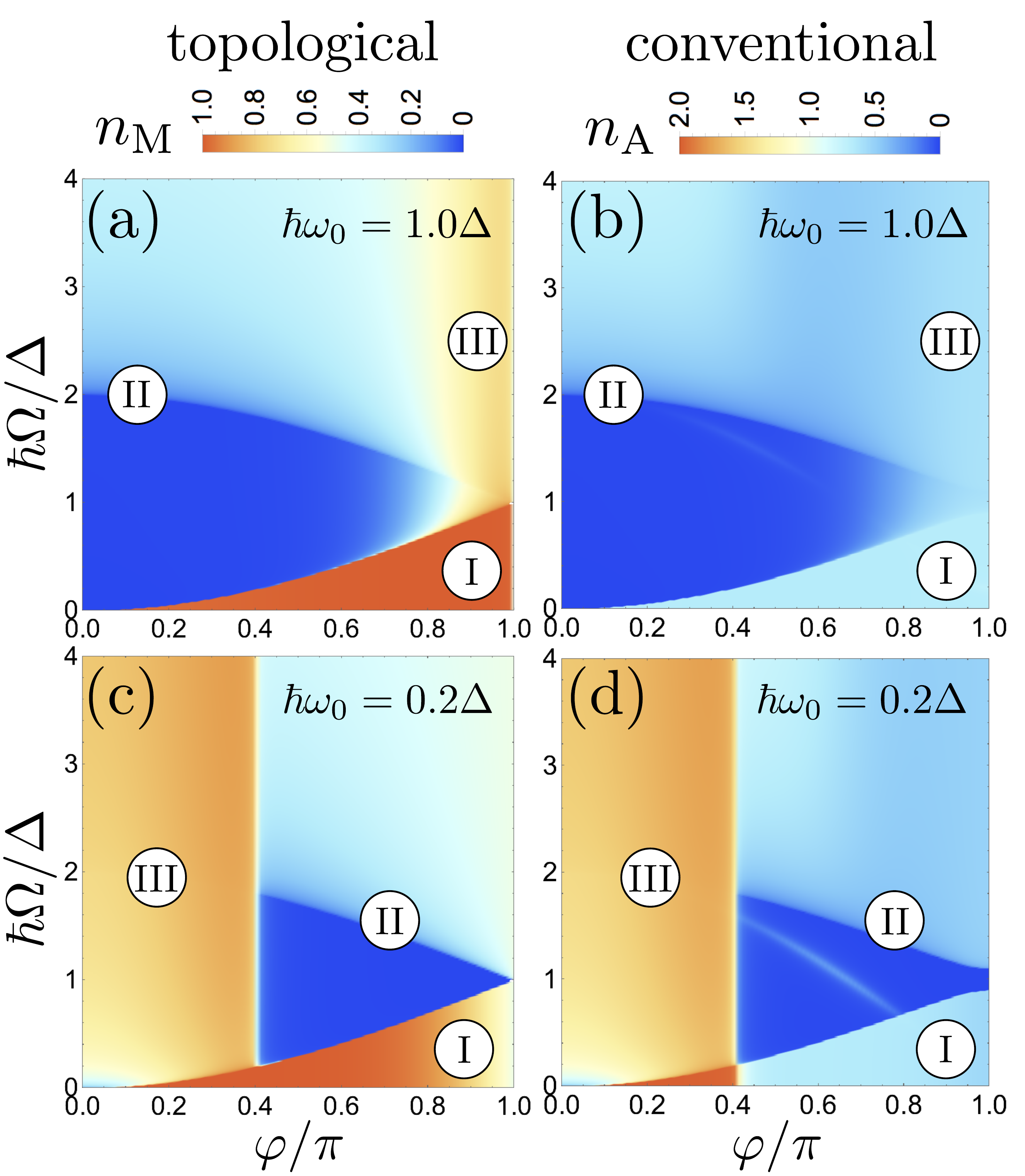}
	\caption{Stationary average bound state occupation in short superconducting junctions as a function of phase difference $\varphi$ and microwave frequency $\Omega$ for different Josephson plasma frequencies $\omega_0$. Majorana bound state occupation $n_\mathrm{M}$ at (a) $\hbar \omega_0 = \Delta$ and (c) $\hbar \omega_0 = 0.2 \Delta$. Andreev bound state occupation $n_\mathrm{A}$ for $\mathcal T = 0.99$ and $\gamma_\mathrm{A} = (2/43)\Delta$ at (b) $\hbar \omega_0 = \Delta$ and (d) $\hbar \omega_0 = 0.2\Delta$. Common parameters: $\gamma = 0.001 \Delta$, $\kb T_\mathrm{qp} = 0.1 \Delta$, $T_\mathrm{env} = 0$ and $(\delta \varphi / \lambda)^2 = 10^{-5}$.}
	\label{fig:occupationMicrowaveConstantResonator}
\end{figure}

Now, we discuss the occupation of the bound states in the presence of microwave radiation.  We plot the occupation as a function of the phase difference $\varphi$ and the microwave frequency $\Omega$.
Considering a setup with a resonator at a fixed energy $\hbar \omega_0 \leq \Delta$, i.e. we focus on the cases $\hbar \omega_0 = 0.2\Delta$ and $\hbar \omega_0 = \Delta$ at $T_\mathrm{env} = 0$, we now allow for transitions due to absorption or emission of energy $\hbar \Omega$ in the presence of a microwave. The resulting occupations are shown in Fig.\,\ref{fig:occupationMicrowaveConstantResonator} where we introduce three regions of interest labeled by (I), (II) and (III).

In region (I), which is given by $\hbar \Omega < \Delta - E_\mathrm{M,A}$, the energy of the microwave is too low for possible transitions of QPs (cf. rates in Eqs.\,(\ref{eqn:MicrowaveEmission}), (\ref{eqn:MicrowaveAbsorption}) and (\ref{eqn:microwaveConventional})). Therefore, the contribution to the rates in this regime is purely due to the
interaction of the SSJ with the damped LC resonator.  
To understand the behavior of the SSJ in region (I), we recall the case without microwave irradiation, shown in 
Fig.\,\ref{fig:occupationResonator}, at $\hbar \omega_0 = 0.2\Delta$ and $\hbar \omega_0 = \Delta$, respectively.
On the one hand, at $\hbar \omega_0 = \Delta$ for the case of the topological junction, 
we have $n_\mathrm{M} \approx 1$ for all phases $\varphi$ whereas, 
for the conventional case at  $\hbar \omega_0 = \Delta$, 
the occupation saturates at $2/3$ (with a switching from $2/3$ to $2$ at a very small phase, 
see discussion in previous section, Fig.\,\ref{fig:occupationResonator}).
On the other hand, both types of junctions show a high occupation at $\hbar \omega_0 = 0.2\Delta$ and phases $\varphi < \varphi_\mathrm{c}$, where the critical phase $\varphi_\mathrm{c}$ is defined by $E_\mathrm{M,A}(\varphi_\mathrm{c}) \approx \Delta - \hbar \omega_0$ which defines the sharp vertical line separating regions (II) and (III). For phases $\varphi > \varphi_\mathrm{c}$, the behaviour differs drastically. In the case of a topological junction, the occupation starts to decrease smoothly from $n_\mathrm{M}(\varphi_\mathrm{c}) \approx 1$ to $n_\mathrm{M}(\pi) = 1/2$, while for the conventional junction the occupation immediately drops from $n_\mathrm{A} \approx 2$ to $n_\mathrm{A}\approx 2/3$ in a short range around $\varphi_\mathrm{c}$.
For $\hbar \Omega > \Delta - E_\mathrm{M,A} $,  microwave photon absorption and emission become possible.

The behaviour of the bound state occupations in region (II) is dominated by transitions due 
to the microwave, i.e. $\Gamma^\mathrm{res}_p \ll \Gamma^\mathrm{mw}_p$, where $p$ labels all possible rates defined in Sec.\,\ref{sec:dynamics}. 
The crossover behavior in region (II) is  set by the condition $\hbar \Omega = \Delta + E_\mathrm{M,A}$
and it can be explained in similar way to the discussion 
for Fig.\,\ref{fig:occupationResonator}
with the difference that below such threshold the absorption processes dominate.
For $\Delta - E_\mathrm{M,A} < \hbar \Omega < \Delta + E_\mathrm{M,A}$, the bound states are almost empty, i.e. $n_\mathrm{M,A} \ll 1$. This is because $\Gamma^\mathrm{mw}_\mathrm{out,1} \gg \Gamma^\mathrm{mw}_\mathrm{in,1}$ due to the small number of continuum QPs at the chosen $T_\mathrm{qp}$. Moreover, the energy of the microwave radiation is still too low for transitions from or into the ground state, i.e. $\Gamma^\mathrm{mw}_\mathrm{out,2} = \Gamma^\mathrm{mw}_\mathrm{in,2} = 0$.
Comparing the topological with the conventional case, there exists a line $\hbar \Omega \approx 2 E_\mathrm{A}$ of nonzero occupation in the case of the conventional junction. This is due to a resonant process associated with the rates $\Gamma^\mathrm{mw}_\mathrm{in/out,AA}$ (cf. Eq.\,(\ref{eqn:AndreevDoubleRateMicrowave})), i.e. $\Gamma^\mathrm{mw}_\mathrm{in/out,AA} \gg \Gamma^\mathrm{mw}_\mathrm{out,1}$, which cannot appear in a topological junction. Away from this resonance, the occupation in the dark blue area is given by $n_\mathrm{M} \approx f(\hbar \Omega + E_\mathrm{M})$ and $n_\mathrm{A} \approx 2f(\hbar \Omega + E_\mathrm{A})$, respectively, with $f(E)$ being the Fermi-Dirac distribution at the temperature $T_\mathrm{qp}$.
At $\hbar \Omega >  \Delta + E_\mathrm{M}$ for the topological junction, the occupation shows almost no phase dependence in region (II) and it is possible to demonstrate that it approaches $n_\mathrm{M}(\hbar \Omega \gg 2\Delta) \approx 0.5$.
At $\hbar \Omega > \Delta + E_\mathrm{A}$ for the conventional junction, there is a weak phase dependence visible due to the different effective density of states. 
However, the occupation also approaches $n_\mathrm{A} \approx 0.5$ in the limit $\hbar \Omega \gg 2\Delta$.
Finally, in Fig.\,\ref{fig:occupationMicrowaveConstantResonator}(c) and Fig.\,\ref{fig:occupationMicrowaveConstantResonator}(d), there is a sharp vertical line separating regions (II) and (III) which is defined by $\hbar \omega_0 = \Delta - E_\mathrm{M,A}$ with $\hbar \omega_0 = 0.2 \Delta$. This line is moved to $\varphi \approx \pi$ since $\hbar \omega_0 = \Delta$ for Fig.\,\ref{fig:occupationMicrowaveConstantResonator}(a) and Fig.\,\ref{fig:occupationMicrowaveConstantResonator}(b). 

In region (III), we have a strong competition between the microwave radiation and the damped resonator. 
At $\hbar \omega_0 = 0.2\Delta$, coupling to the resonator leads to a refilling of the bound state while the microwave contribution leads to a depopulation. This results in occupations $0.5 < n_\mathrm{M} < 1$ and $1 < n_\mathrm{A} < 2$ for the topological and the conventional junction, repsectively.
For $\hbar \omega_0 = \Delta$, the resonator  also leads to a strong refilling of the bound state in the case of the topological junction with $0.5 < n_\mathrm{M} < 1$. For the conventional junction, this effect is much weaker  and 
the ABS  is still  unoccupied, i.e.  $ n_\mathrm{A}  < 1$.

\begin{figure}
	\centering
	\includegraphics[width=0.9\linewidth]{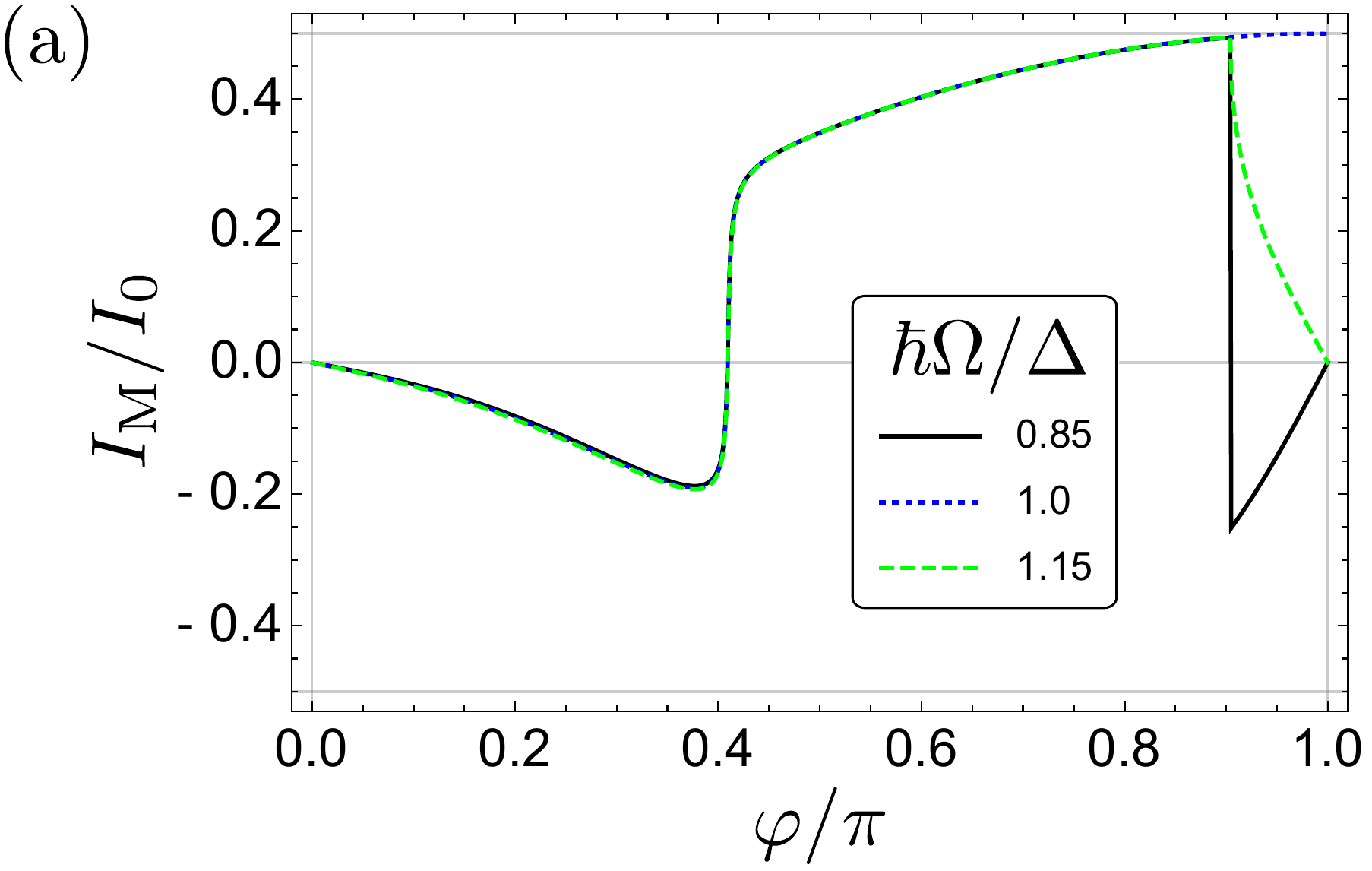}
	\includegraphics[width=0.9\linewidth]{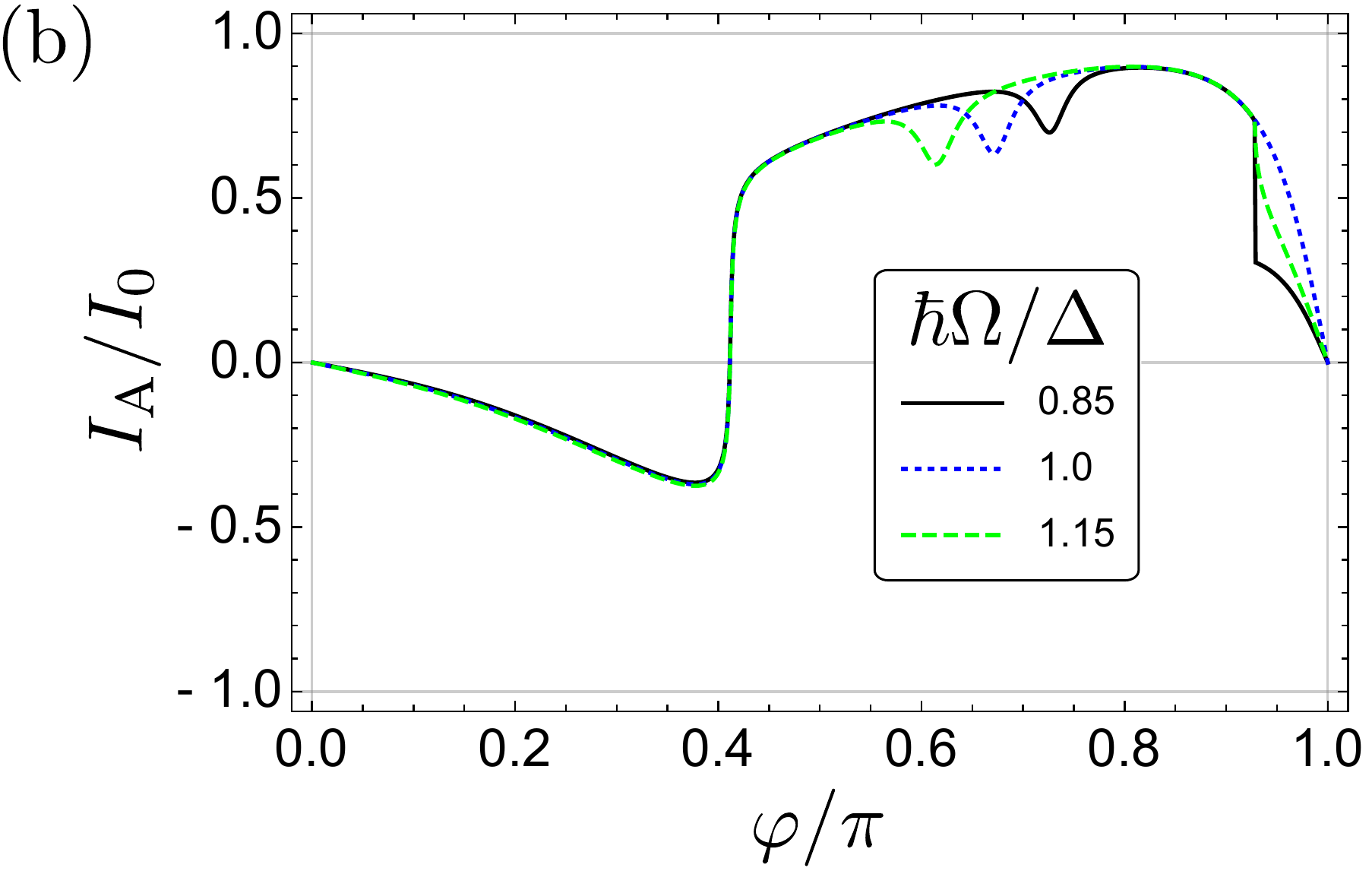}
	\caption{Supercurrent as a function of phase difference $\varphi$ for different microwave frequencies $\Omega$ in the presence of the resonator with plasma frequency $\omega_0 = 0.2 \Delta/\hbar $. (a) Current $I_\mathrm{M}$ carried by Majorana bound states in short topological superconducting junctions. (b) Current $I_\mathrm{A}$ carried by Andreev bound states in short conventional superconducting junctions. Parameters are the same as for Fig.\,\ref{fig:occupationMicrowaveConstantResonator}(c) and (d), respectively. Current is plotted in units of $I_0 = e \Delta / \hbar$.}
	\label{fig:CurrentWithMicrowave}
\end{figure}

With the occupations of the bound states given in Fig.\,\ref{fig:occupationMicrowaveConstantResonator}, we calculate the corresponding currents carried by these states. 
The current is shown in Fig.\,\ref{fig:CurrentWithMicrowave} at $\hbar \omega_0 = 0.2\Delta$ for different microwave frequencies $\Omega$.
For the case of a topological junction, plotted in Fig.\,\ref{fig:CurrentWithMicrowave}(a), the current shows two zero crossing
points  for $\hbar \Omega < \Delta$, while there is only a single crossing left at $\hbar \Omega > \Delta$.
 As already discussed for the occupation,  the first zero crossing appears at a phase $\varphi$ for which $E_\mathrm{M}(\varphi) \approx \Delta - \hbar \omega_0$, 
 while the second zero current crossing at  $\hbar \Omega < \Delta$ 
 takes place at   $E_\mathrm{M}(\varphi) \approx \Delta - \hbar \Omega$.
 In a similar way, for $\hbar \Omega > \Delta$, we have a kink  at $E_\mathrm{M}(\varphi) \approx \hbar \Omega - \Delta$.
For the conventional junction plotted in Fig.\,\ref{fig:CurrentWithMicrowave}(b), 
we observe only one zero crossing for all values of $\Omega$. 
The origin is similar to the case of the topological junction, i.e. it appears at 
$E_\mathrm{A} \approx \Delta - \hbar \omega_0$. 
Moreover, there are dips in the current phase relation which correspond to  transitions of a Cooper pairs between the ground state and the excited state at $2E_\mathrm{A} = \hbar \Omega$, as discussed in  Ref. [\onlinecite{Bergeret:2010bw}].

%
%
%
\section{Summary and conclusions}
\label{sec:conclusion}
We calculated the nonequilibrium bound state occupations for short topological and short conventional superconducting junctions being part of a dc-SQUID in the presence of an applied ac microwave field at frequency $\Omega$ and in the presence of phase fluctuations due to a damped resonator 
at frequency $\omega_0$.
We used a simple rate equation to obtain the stationary state of the occupations. 
We assumed that the continuum quasiparticles above the gap relax much faster than the dynamics of the 
discrete ABS level and we discussed the case when quasiparticles are still described  by a Fermi distribution 
but with higher effective temperature 
$T_\mathrm{qp} > T_\mathrm{env}$ with respect to the environmental temperature in the system. 
This mimics, in a phenomenological way, an effective nonequilibrium distribution.

In the absence of a microwave field, at low  environmental temperature $(\kb T_\mathrm{env} \ll \hbar \omega_0)$, 
we have shown that if the resonator's Josephson plasma energy is $\hbar \omega_0 < 2 \Delta$, 
the resulting occupations of the short superconducting junctions differ drastically for the topological and conventional case (cf. Fig.\,\ref{fig:occupationResonator}) despite the fact that the transmission is $\mathcal{T} \approx 1$ in the case of the conventional junction.
This result is due to the process of  transfer a Cooper pair between the ground state and the 
excited bound state --- occurring with energy $2 E_\mathrm{A}$ --- in the conventional junction that leads to a decreased occupation.
In contrast, the lack of this process in the topological case  leads to a high occupation 
in a wide region.
In the presence of photon pumping due to microwaves, the occupations can be generally 
determined by the competition of microwaves and the photon emitted in the damped resonator.
If the microwave dominates, we found that the steady occupation of the bound states 
is in correspondence of the quasiparticle temperature (blue regions in Fig.\,\ref{fig:occupationMicrowaveConstantResonator}).
Hence, in this regime of strong pumping,  measuring the equilibrium occupation can give a priori information about the 
nonequilibrium quasiparticle population.
Finally, we have shown that the regions of different occupation and their switching and crossover 
appear in a one to one correspondence in the behavior of the supercurrent as a function of the phase difference.

Before to conclude, we further discuss possible methods to measure the nonequilibrium population.
In experiments for the SAC \cite{Bretheau:2013by,Bretheau:2013bt} as well as in NW \cite{vanWoerkom:2017gl},
the SQUID system sketched in Fig.\,\ref{fig:model} was capacitively coupled to a voltage 
biased Josephson junction acting as an (incoherent) emitter of photons directed towards 
the SQUID as well as a spectrometer of the system. 
The  external Josephson junction of the spectrometer was designed with a typical  characteristic impedance $\mathrm{Re}[Z_0] \ll R_\mathrm{Q}$, with $R_\mathrm{Q} = h / 4 e^2$, such that the current observed at finite voltage bias $V_\mathrm{spec}$ can be expressed, using $P(E)$ theory
\cite{Ingold:92},
as a function of the impedance seen by the junctions itself \cite{Holst:1994dh,Hofheinz:2011jc}
\begin{equation}
I_\mathrm{spec} = \frac{I^2_\mathrm{c,spec}}{2} \, \frac{\mathrm{Re}[Z(\omega)]}{V_\mathrm{spec}} \, 
\end{equation}
in which $Z(\omega)$ corresponds to the impedance associated with the SQUID and $I_\mathrm{c,spec}$ is the critical current of the spectrometer.
Far away from the frequency range associated with the plasma mode $\omega_0$, one can reasonably assume
that the impedance of the SQUID is essentially given by the impedance of the SSJ, i.e.
$Z(\omega) \approx Z_\mathrm{SSJ} (\omega)$. 
The latter quantity is the result of the  processes of emission and absorption occurring in 
the SSJ  (described in this work) and it is directly related to the occupation numbers of the ABS in the conventional SSJ \cite{Kos:2013hl} as well as in the topological one \cite{Vayrynen:2015ga,Peng:2016bf}. 

Another way to measure the ABS occupation can be via switching current measurements\cite{Zgirski:2011dx}.
By measuring the switching current, these occupations should be experimentally accessible since the supercurrent carried by the bound states is proportional to the occupation, as discussed in the theoretical work of Ref.\,[\onlinecite{Peng:2016bf}].

\acknowledgments 
We thank L. I. Glazman, J. I. V{\"a}yrynen and G. Catelani for discussions.
This work was supported by the Excellence Initiative through the Zukunftskolleg 
and the YSF Fund of the University of Konstanz and by the DFG
through the SFB 767 and grant RA 2810/1-1.

\appendix
%
%
%
\section{Derivation of the interacting Hamiltonian}
\label{appendix:c}
In this appendix, we provide the derivation of the final Hamiltonian of the combined SQUID system given in Eq.\,(\ref{eqn:totalHamiltonian}) consisting of the short superconducting junction (SSJ), the damped LC resonator and the microwave radiation due to a small ac part of the magnetic flux penetrating the SQUID.

\subsection{Hamiltonian of the LC resonator}
\label{appendix:c1}
First, we start with the Hamiltonian of the damped LC resonator (see Sec.\,\ref{subsec:resonator}) which is formed by a conventional Josephson junction in the Josephson regime, together with dissipation described within the Caldeira-Leggett model via coupling to a bath. The general Hamiltonian of a Josephson junction is given by
\begin{align}
	H_\mathrm{res} = E_\mathrm{C} \left( N - \frac{Q_\mathrm{r}}{2e} \right)^2 - E_\mathrm{J} \cos\chi \, ,
\end{align}
with the offset charge $Q_\mathrm{r} = e N_\mathrm{r}$ carried by $N_\mathrm{r}$ single charge carriers, the number of charge carrying Cooper pairs $N$ and the phase difference $\chi$ across the junction which satisfy the commutation relation $[\chi,N]=\i$. In the Josephson regime for $E_\mathrm{J} \gg E_\mathrm{C}$, phase fluctuations are small, i.e. $\left\langle\chi^2\right\rangle \ll 1$. Moreover, we write the accumulated charge in a symmetrized way, $Q_\mathrm{r} = (e/2) \, \delta n$, in terms of the quasiparticle operators of the SSJ, with
\begin{align}
	\delta n
	= 
	\left\{\begin{array}{l c l}
		\sum_{\sigma} \int \d x \, \sgn(x) \, \psi_\sigma^\dag \psi_\sigma^{\phantom{\dag}} & , & \mathrm{topological}
		\\
		\\
		\sum_{\alpha\sigma} \int \! \d x \, \sgn(x) \, \psi_{\alpha\sigma}^\dag \psi_{\alpha\sigma}^{\phantom{\dag}} & , & \mathrm{conventional} 
	\end{array}	\right. 
	\label{eqn:difference}
\end{align}
being the difference in number of particles between the right and left leads, 
spin $\sigma = \, \uparrow,\downarrow$ and $\alpha = \mathrm{R,L}$ referring to right/left-movers.
Expanding the resulting Hamiltonian to lowest order in $\chi$, we obtain
\begin{align}
	H_\mathrm{res} = E_\mathrm{C} \left( N - \frac{\delta n}{4}  \right)^2 + \frac{E_\mathrm{J}}{2} \chi^2 \, .
\end{align}
By means of the unitary transformation $U = \exp(-\i \, \chi \, \delta n/4)$, we achieve the shift $ N \to N + \delta n/4 $ yielding the final result presented in Eq.\,(\ref{eqn:HamiltonianResonator}) (but without the bath), where we have introduced the Josephson plasma frequency $\omega_0 = \sqrt{2 E_\mathrm{C} E_\mathrm{J}}/\hbar$ together with bosonic creation and annihilation operators $b_0^\dag$ and $b_0^{\phantom{\dag}}$, respectively, satisfying $\chi = \sqrt{E_\mathrm{C} / \hbar \omega_0} \, (b_0^\dag + b_0^{\phantom{\dag}})$ and $N = (\i/2) \sqrt{\hbar \omega_0/E_\mathrm{C}} \, (b_0^\dag - b_0^{\phantom{\dag}})$.

\subsection{Hamiltonian of the short superconducting junction coupled to the LC resonator}
\label{appendix:c2}
Now, we derive the final Hamiltonian of the SSJ (see Sec.\,\ref{subsec:ssj}) and the interactions (see Sec.\,\ref{subsec:interaction}). We start from
\begin{align}
	H_\mathrm{SSJ}^{(\beta)} = \frac{1}{2} \int \! \d x \, \Psi^\dag_{\beta}(x) \, \mathcal H_{\beta}(x) \, \Psi_{\beta}^{\phantom{\dag}}(x) \, ,
\end{align}
with the Hamiltonian for the short topological junction ($\beta = \mathrm{tj}$) in the Fu-Kane \cite{Fu:2008gu} model
\begin{equation}
\mathcal H_\tj(x) = - \i \hbar v_\tj \sigma_3 \tau_3 \partial_x - \mu \tau_3 + \Delta(x) \, e^{\i\phi(x)\tau_3}  \tau_1
\label{eqn:hamAppTJ}
\end{equation}
and in the short conventional junction\cite{nabl09} ($\beta = \mathrm{cj}$)
\begin{multline}
\mathcal H_\cj(x) = - \i \hbar v_\cj  \sigma_3 \tau_3 \partial_x 
\\
+ \hbar v_\cj  Z \, \delta(x) \, \sigma_1 \tau_3 + \Delta(x) \, e^{\i\phi(x)\tau_3}  \tau_1 \, ,
\label{eqn:hamAppCJ}
\end{multline}
both with the inhomogeneous pairing potential $\Delta(x)$ and the phase difference $\phi(x)$ which are given by Eq.\,(\ref{ch4:phaseTopJunc}) in the short junction limit $L \to 0$.
Since the SSJ is placed in the dc-SQUID, all phases are related according to Eq.\,(\ref{eqn:link}), i.e. we replace the phase $\phi \to \varphi + \chi$ in $\phi(x)$ (the time-dependence will be discussed in appendix \ref{appendix:c3}). The expressions of the field operators $\Psi_{\beta}^{\phantom{\dag}}$ are given in Sec. \ref{subsec:ssj}.

In a first step, we expand the Hamiltonian of the SSJ as $H_\mathrm{SSJ}^{(\beta)}(\phi) \approx H_\mathrm{SSJ}^{(\beta)}(\varphi) + \chi J_0$ to linear order in $\chi$ with the operator $J_0 = \frac{\partial H_\mathrm{SSJ}^{(\beta)}(\phi) }{\partial \chi} |_{\chi=0}$. 
Now we apply the unitary transformation $U = \exp(-\i \, \chi \, \delta n / 4) \approx 1 - \frac{\i}{4} \chi \, \delta n$ that we have used for the Hamiltonian of the resonator (see appendix \ref{appendix:c1}), such that the linearized transformation reads
\begin{align}
	\tilde{H}_\mathrm{SSJ}^{(\beta)} 
	&= U H_\mathrm{SSJ}^{(\beta)}(\phi) U^\dag 
	= U H_\mathrm{SSJ}^{(\beta)}(\varphi) U^\dag + U \chi J_0 U^\dag \nonumber \\
	&\approx H_\mathrm{SSJ}^{(\beta)}(\varphi) + \left( \frac{\i}{4} \chi \left[H_\mathrm{SSJ}^{(\beta)}(\varphi) , \delta n \right] \right) + \chi J_0 \nonumber \\
	&= H_\mathrm{SSJ}^{(\beta)}(\varphi) + \left(  \chi \Phi_0 I_\beta - \chi J_0  \right) + \chi J_0 \nonumber \\
	&= H_\mathrm{SSJ}^{(\beta)}(\varphi) +   \chi \Phi_0 I_\beta
	\label{eq:transformedHamiltonian}
\end{align}
with the current operator $I_\beta$ as defined in Eq.\,(\ref{eqn:currentOperatorSSJ}) and the flux quantum $\Phi_0 = \hbar / 2e$. Now, the interaction Hamiltonian between the resonator and the SSJ is the second term in Eq.\,(\ref{eq:transformedHamiltonian}) which we define as
\begin{align}
	H_\mathrm{int}^{(\beta)} = \lambda \Phi_0 (b_0^\dag + b_0^{\phantom{\dag}}) I_\beta
\end{align}
by replacing $\chi$ as explained in appendix \ref{appendix:c1}. This interaction describes the coupling of the SSJ to the resonator and is presented in Eq.\,(\ref{eqn:interactionRES}) in the main text.

\subsection{Hamiltonian of the short superconducting junction coupled to a microwave field}
\label{appendix:c3}

Now, we replace $\varphi \to \varphi + \delta\varphi(t)$ in Eq.\,(\ref{eq:transformedHamiltonian}), taking a small time-dependent modulation of the magnetic flux into account. By means of the local and time-dependent unitary transformation $U(x,t) = \exp(-\i \, \sgn(x) \, \delta\varphi(t) \, \tau_3 / 4)$ which neither affects the Hamiltonian of the resonator $H_\mathrm{res}$ nor the interaction $H_\mathrm{int}^{(\beta)}$, we remove the time-dependence $\delta\varphi(t)$ from $H_\mathrm{SSJ}^{(\beta)}(\varphi + \delta\varphi(t))$. The transformation of the time-dependent BdG equation results in
\begin{multline}
	U(x,t) \bigl[ \mathcal H_\beta(\varphi + \delta\varphi(t)) - \i \hbar \partial_t \bigr] U^\dag(x,t) = \mathcal H_\beta(\varphi) 
	 \\
	  - \i \hbar \partial_t  +  \frac{e V(t)}{2} \sgn(x)  \tau_3 + \frac{\hbar v_\beta}{2} \sigma_3 \, \delta(x) \, \delta\varphi(t)   \, ,
	 \label{eqn:firstTrafo}
\end{multline}
creating two additional contributions. The time-dependent voltage $V(t)$ is given by the Josephson relation $V(t) = \Phi_0 \, \partial_t \, \delta\varphi(t)$ whose contribution will be neglected in the following\footnote{Notice that we work in a different gauge as compared to the work Ref.\,[\onlinecite{Zazunov:2005ec}].}. Transition rates are part of the real part of the admittance while a time-dependent voltage leads to effects in the imaginary part in the admittance and thus does not affect transition rates\cite{Kos:2013hl}. The last term in Eq.\,(\ref{eqn:firstTrafo}) gives our second contribution to the interaction Hamiltonian which we define as
\begin{align}
	H_\mathrm{mw}^{(\beta)}(t) = \delta\varphi(t) \, \Phi_0 \, I_\beta \, ,
\end{align}
with the current operator as defined in Eq.\,(\ref{eqn:currentOperatorSSJ}). 
This is the interaction presented in Eq.\,(\ref{eqn:interactionMW}) in the main text describing the coupling of the SSJ to the classical microwave field $\delta\varphi(t)$,

%
%
%
\section{Solutions of the Bogoliubov-de Gennes equations}
\label{appendix:a}
In this appendix, we provide the solutions of the Bogoliubov-de Gennes (BdG) equation for the short topological (appendix \ref{subappendix:a1}) and the short conventional (appendix \ref{subappendix:a2}) superconducting junction.

\subsection{Solutions for the topological junction}
\label{subappendix:a1}
The BdG Hamiltonian for the topological junction in the short junction limit is given in Eq.\,(\ref{ch4:bdgHamiltonianTopologicalJunction}). Applying the local unitary transformation $U(x) = e^{-\i \varphi \, \sgn(x) \tau_3 / 4}$ on this Hamiltonian, $\mathcal H_\tj(x) \to \mathcal H_\tj'(x) = U(x) \mathcal H_\tj(x) U^\dag(x) $, we obtain
\begin{multline}
	\mathcal H_\tj'(x) = - \i \hbar v_\tj \sigma_3 \tau_3 \partial_x - \mu \, \tau_3 \\
				+ \Delta(x) \,  \tau_1 + \hbar v_\tj  \frac{\varphi}{2} \,  \sigma_3 \,  \delta(x) \, ,
\end{multline}
from which we find the solutions to the BdG equation $\mathcal H_\tj'(x) \Phi_E(x) = E \Phi_E(x)$.
Since this is a first-order differential equation, we write the solutions as $\Phi_E(x) = B(x,x_0) \Phi_E(x_0)$ with the transfer matrix 
\begin{multline}
B(x,x_0) 
= 
\mathrm{P} \exp\biggl( \i \frac{\tau_3\sigma_3}{\hbar v_\tj} \int_{x_0}^{x} \!\!\! \d x' \, 
\Bigl[ E - \Delta(x') \tau_1 
\\
+ \mu \,  \tau_3 - \hbar v_\tj  \frac{\varphi}{2}  \sigma_3  \, \delta(x') 
\Bigr] \biggr)
\label{eq:boundaryTopJunc}
\end{multline}
for a reference point $x_0 < x$. Hereby we introduced the ordering operator $\mathrm{P}$ which moves larger $x$ to the left. Hence, the boundary condition at the interface $x=0$ is given by
\begin{equation}
\lim\limits_{\varepsilon \to 0^+}B(\varepsilon , -\varepsilon ) = B(0^+,0^-) = e^{- \i \varphi \tau_3 / 2} 
\label{eqn:boundaryTJ}
\end{equation}
which links the solutions on the left and right side of the interface. Due to particle-hole symmetry described in Sec.\,\ref{subsubsec:topological}, we can restrict the calculation to positive energies $E \geq 0$.

\subsubsection{Continuum wave functions}
\label{subappendix:a1a}
For $E > \Delta$, we calculate scattering states $\Phi_{E,s}(x)$, $s \in \{1,2,3,4\}$, which correspond to four possible incident quasiparticles. First of all, each incident quasiparticle creates four outgoing quasiparticles
\begin{subequations}
	\begin{align}
	\Phi_\leftarrow^\mathrm{h} (x) &= \sqrt{\frac{\mathcal N_E}{\mathcal L}} \bigl( e^{-\alpha/2} ,  e^{\alpha/2} , 0 ,  0 \bigr)^\mathrm{T} e^{\i k_\mathrm{h} x} \, \Theta(-x) \, ,\\
	\Phi_\leftarrow^\mathrm{e} (x) &= \sqrt{\frac{\mathcal N_E}{\mathcal L}} \bigl(  0 , 0 ,  e^{\alpha/2} ,  e^{-\alpha/2} \bigr)^\mathrm{T} e^{-\i k_\mathrm{e} x} \, \Theta(-x) \, , \\
	\Phi_\rightarrow^\mathrm{e} (x) &= \sqrt{\frac{\mathcal N_E}{\mathcal L}} \bigl( e^{\alpha/2} , e^{-\alpha/2}  , 0 ,  0 \bigr)^\mathrm{T} e^{\i k_\mathrm{e} x} \, \Theta(x) \, ,\\
	\Phi_\rightarrow^\mathrm{h} (x) &= \sqrt{\frac{\mathcal N_E}{\mathcal L}} \bigl(  0 ,   0 , e^{-\alpha/2} ,  e^{\alpha/2} \bigr)^\mathrm{T} e^{-\i k_\mathrm{h} x} \, \Theta(x) \, , 
	\end{align}
	\label{eqn:outgoingTJ}
\end{subequations}
where $\Phi_q^r$ is a $r$-like (e: electron, h: hole) quasiparticle moving to the $q$ ($\leftarrow$: left, $\rightarrow$: right) lead. Here, we defined the energy dependent scattering phase $\alpha(E)$ via
\begin{align}
e^{\pm \alpha(E)} = \frac{E}{\Delta} \pm \sqrt{\frac{E^2}{\Delta^2} - 1} \, ,
\label{eqn:energyScattering}
\end{align}
and the wave numbers $k_\mathrm{e,h}(E) = k_\tj \pm \kappa(E)$ for electron- and hole-like quasiparticles, with the definitions $k_\tj = \mu / \hbar v_\tj$ and 
\begin{equation}
	\kappa(E) = \frac{\sqrt{E^2 - \Delta^2}}{\hbar v_\tj} \, .
\end{equation}
Moreover, $\mathcal N_E = \Delta/2E$ is a normalization constant and $\mathcal L$ is a length-scale over which the propagating quasiparticle states are defined.

In the same way, we define the four possible incident quasiparticles as
\begin{subequations}
	\begin{align}
	\Phi_1^\mathrm{in} (x) &= \sqrt{\frac{\mathcal N_E}{\mathcal L}} \bigl( e^{\alpha/2} , e^{-\alpha/2} , 0  , 0 \bigr)^\mathrm{T} e^{\i k_\mathrm{e} x} \, \Theta(-x) \, ,\\
	\Phi_2^\mathrm{in} (x) &= \sqrt{\frac{\mathcal N_E}{\mathcal L}} \bigl(  0  , 0 , e^{-\alpha/2} ,  e^{\alpha/2} \bigr)^\mathrm{T} e^{-\i k_\mathrm{h} x} \, \Theta(-x) \, , \\
	\Phi_3^\mathrm{in} (x) &= \sqrt{\frac{\mathcal N_E}{\mathcal L}} \bigl( 0 ,   0 , e^{\alpha/2} , e^{-\alpha/2} \bigr)^\mathrm{T} e^{-\i k_\mathrm{e} x} \, \Theta(x) \, ,\\
	\Phi_4^\mathrm{in} (x) &= \sqrt{\frac{\mathcal N_E}{\mathcal L}} \bigl(  e^{-\alpha/2} ,  e^{\alpha/2} , 0 , 0 \bigr)^\mathrm{T} e^{\i k_\mathrm{h} x} \, \Theta(x) \, .
	\end{align}
	\label{eqn:incidentTJ}
\end{subequations}
With the incident and outgoing quasiparticles defined in Eqs.\,(\ref{eqn:incidentTJ}) and (\ref{eqn:outgoingTJ}), respectively, we write the four scattering states as
\begin{multline}
	\Phi_{E,s}(x) =  \Phi_s^\mathrm{in}(x) +  A_s \Phi_\leftarrow^\mathrm{h}(x) +  B_s \Phi_\leftarrow^\mathrm{e}(x) \\
	+ C_s \Phi_\rightarrow^\mathrm{e}(x) + D_s \Phi_\rightarrow^\mathrm{h}(x) \, .
	\label{eqn:scatteringStatesTJ}
\end{multline}

The corresponding scattering coefficients $A_s$, $B_s$, $C_s$ and $D_s$ are obtained by using the boundary condition (\ref{eqn:boundaryTJ}) and solving the equation
\begin{equation}
	\Phi_{E,s}(0^+) = B(0^+,0^-) \, \Phi_{E,s}(0^-)
\end{equation}
for each of the four scattering states. The full solution reads
\begin{subequations}
	\begin{align}
	\bigl(
		A_1 , B_1 , C_1 , D_1 
	\bigr)
	&=
	\bigl(
		\mathcal A , 0 , \mathcal B , 0
	\bigr)
	\, , \\
	\bigl(
		A_2 , B_2 , C_2 , D_2 
	\bigr)
	&=
	\bigl(
		0 , \mathcal A^* , 0 , \mathcal B^*
	\bigr)
	\, , \\
	\bigl(
		A_3 , B_3 , C_3 , D_3
	\bigr)
	&=
	\bigl(
		0 , \mathcal B^* , 0 , \mathcal A^*
	\bigr)
	\, , \\
	\bigl(
		A_4 , B_4 , C_4 , D_4
	\bigr)
	&=
	\bigl(
		\mathcal B , 0 , \mathcal A , 0 
	\bigr) \, , 
	\end{align}
\end{subequations}
with the scattering coefficients
\begin{subequations}
	\begin{align}
	\mathcal A &= \frac{\Delta^2}{E^2 - E_\mathrm{M}^2} \, (- \i ) \, \sin\frac{\varphi}{2} \, \sinh\Bigl( \alpha - \i \frac{\varphi}{2} \Bigr) \, ,  \\
	\mathcal B &= \frac{\Delta^2}{E^2 - E_\mathrm{M}^2}  \, \sinh \alpha  \, \sinh\Bigl( \alpha - \i \frac{\varphi}{2} \Bigr) \, ,
	\end{align}
\end{subequations}
satisfying $|\mathcal A|^2 + |\mathcal B|^2 = 1$. Here, we already introduced $E_\mathrm{M} = \Delta \cos(\varphi/2)$ which is the energy of the Majorana bound state inside the gap (see appendix \ref{subappendix:a1b}).

\subsubsection{Majorana bound state wave function}
\label{subappendix:a1b}
For energies $E<\Delta$, there is no incident quasiparticle and $\alpha \to \i \alpha$ and $\kappa \to \i \kappa$ become complex. Therefore, the scattering state for the Majorana bound state following from Eq.\,(\ref{eqn:scatteringStatesTJ}) reads
\begin{multline}
	\Phi_\mathrm{M}(x) 
	= 
	\begin{pmatrix} 
		e^{-\i \alpha / 2} A_0 \exp(\i k_\tj x )  \\ 
		e^{\i \alpha / 2} A_0 \exp(\i k_\tj x ) \\ 
		e^{\i \alpha / 2} B_0 \exp(-\i k_\tj x ) \\ 
		e^{-\i \alpha / 2} B_0 \exp(- \i k_\tj x )
	\end{pmatrix} e^{\kappa x} \, \Theta(-x) 
	\\
	+ 
	\begin{pmatrix} 
		e^{\i \alpha / 2} C_0 \exp(\i k_\tj x )  \\ 
		e^{-\i \alpha / 2} C_0 \exp(\i k_\tj x ) \\ 
		e^{-\i \alpha / 2} D_0 \exp(-\i k_\tj x ) \\ 
		e^{\i \alpha / 2} D_0 \exp(- \i k_\tj x )
	\end{pmatrix} e^{- \kappa x} \, \Theta(x) 
	\label{eq:mbsWF}
\end{multline}
with $\alpha(E)$ defined via
\begin{align}
	e^{\pm \i \alpha(E)} = \frac{E}{\Delta} \pm \i \sqrt{ 1 - \frac{E^2}{\Delta^2} } \, .
	\label{eqn:energyScatteringTJ}
\end{align}
and the wave numbers $k_\tj = \mu / \hbar v_\tj$ and
\begin{equation}
	\kappa(E)  = \frac{\sqrt{\Delta^2 - E^2}}{\hbar v_\tj}  \, .
\end{equation}
Applying the boundary condition in Eq.\,(\ref{eqn:boundaryTJ}), i.e.
\begin{equation}
	\Phi_\mathrm{M}(0^+) = B(0^+,0^-) \, \Phi_\mathrm{M}(0^-) \, ,
	\label{eqn:boundaryMajorana}
\end{equation}
we find the energy has to fulfill $E = E_\mathrm{M}$ with the Majorana bound state energy
\begin{equation}
	E_\mathrm{M}(\varphi) = \Delta \cos\frac{\varphi}{2} \, ,
\end{equation}
in order to find nontrivial solutions for the scattering coefficients $A_0$, $B_0$, $C_0$ and $D_0$. Under this condition, Eq.\,(\ref{eqn:boundaryMajorana}) reveals
\begin{subequations}
	\begin{align}
	A_0 &= C_0 \, , \\
	B_0 &= D_0 \, .
	\end{align}
\end{subequations}
To normalize the subgap wave function, we use
\begin{equation}
1 = \int_{-\infty}^\infty \!\! \! \d x \, \bigl( \Phi_\mathrm{M}^*(x) \bigr)^\mathrm{T} \Phi_\mathrm{M}(x) = \frac{2}{\kappa_\mathrm{M}} \Bigl( |A_0|^2 + |B_0|^2 \Bigr) \, ,
\end{equation}
with $\kappa_\mathrm{M} = \kappa(E_\mathrm{M})$. In the absence of a magnetic field in the nonsuperconducting part of the topological junction, we lack of one more condition for the scattering amplitudes since left-/right-moving quasiparticles experience no normal scattering without a magnetic field due to the helicity of the conducting states. To match the solution found in Ref.\,[\onlinecite{Peng:2016bf}] in this limit, we take $A_0 = 0$. Finally, we obtain the solution
\begin{subequations}
	\begin{align}
		A_0 &= 0 \, , \\
		B_0 &= \sqrt{\frac{\Delta}{2\hbar v_\tj} \sin\frac{\varphi}{2}} \, .
	\end{align}
\end{subequations}
Using the MBS wave function given in Eq.\,(\ref{eq:mbsWF}) and the definitions of the BdG QP operators
\begin{subequations}
\begin{align}
	\gamma_{E_\mathrm{M}}^{\phantom{\dag}} &= \int_{-\infty}^\infty \!\! \! \d x \, \bigl( \Phi_\mathrm{M}^*(x) \bigr)^\mathrm{T} \,  \Psi_{\tj}^{\phantom{\dag}}(x) \, , \\
	 \gamma_{-E_\mathrm{M}}^\dag &= \int_{-\infty}^\infty \!\! \! \d x \,  \Psi_{\tj}^\dag(x) \,  \Phi_\mathrm{M}^{\phantom{\dag}}(x) \, ,
\end{align}
\end{subequations}
one can show the Majorana property $\gamma_0^{\phantom{\dag}} = \gamma_0^\dag$ of the BdG QP operators in the limit $\varphi \to \pi$ and $E_\mathrm{M} \to 0$.

\subsection{Solutions for the conventional junction}
\label{subappendix:a2}
The BdG Hamiltonian for the conventional junction in the short junction limit is given in Eq.\,(\ref{ch4:bdgHamiltonianConventionalJunction}). Applying the local unitary transformation $U(x) = e^{-\i \varphi \, \sgn(x) \tau_3 / 4}$ on this Hamiltonian, $\mathcal H_\cj(x) \to \mathcal H_\cj'(x) = U(x) \mathcal H_\cj(x) U^\dag(x) $, we obtain
\begin{multline}
 \mathcal H_\cj'(x) = - \i \hbar v_\cj  \sigma_3 \tau_3 \partial_x + \Delta(x) \,  \tau_1
	\\
	+ \hbar v_\cj  \biggl[ \frac{\varphi}{2}  \sigma_3  +  Z  \, \sigma_1 \tau_3  \biggr] \delta(x)
\end{multline}
from which we find the solutions to the BdG equation $\mathcal H_\cj'(x) \Phi_E(x) = E \Phi_E(x)$.
Since this is a first-order differential equation, we write the solutions as $\Phi_E(x) = B(x,x_0) \Phi_E(x_0)$ with the transfer matrix 
\begin{multline}
	B(x,x_0) 
	= 
	\mathrm{P} \exp\biggl( \i \frac{\sigma_3 \tau_3}{\hbar v_\cj} \int_{x_0}^{x} \!\!\! \d x' \, 
	\Bigl[ E - \Delta(x') \tau_1 
	\\
	- \hbar v_\cj  \Bigl( \frac{\varphi}{2}  \sigma_3  +  Z  \, \sigma_1 \tau_3  \Bigr) \delta(x') 
	\Bigr] \biggr)
\end{multline}
for a reference point $x_0 < x$. Again, P denotes the ordering operator as introduced in Eq.\,(\ref{eq:boundaryTopJunc}). Hence, the boundary condition at the interface $x=0$ is given by
\begin{align}
	\lim\limits_{\varepsilon \to 0^+}B(\varepsilon , -\varepsilon ) &= B(0^+,0^-) \nonumber \\
	&= e^{- \i \varphi \tau_3 / 2} \frac{1}{\sqrt{\mathcal T}} \Bigl( 1 + \sigma_2 \sqrt{1-\mathcal T} \Bigr)
	\label{eqn:boundary}
\end{align}
which links the solutions on the left and right side of the interface. In Eq.\,(\ref{eqn:boundary}), we have defined the transmission $\mathcal T = 1/\cosh^2 Z$. Due to particle-hole symmetry described in Sec.\,\ref{subsubsec:conventional}, we can restrict the calculation to positive energies $E \geq 0$.

\subsubsection{Continuum wave functions}
\label{subappendix:a2a}
For $E > \Delta$, we calculate scattering states $\Phi_{E,s}(x)$, $s \in \{1,2,3,4\}$, which correspond to four possible incident quasiparticles. First, we define the four possible outgoing quasiparticles
\begin{subequations}
	\begin{align}
		\Phi_\leftarrow^\mathrm{h} (x) &= \sqrt{\frac{\mathcal N_E}{\mathcal L}} \bigl( e^{-\alpha/2} ,  e^{\alpha/2} , 0 , 0 \bigr)^\mathrm{T} e^{-\i k x} \, \Theta(-x) \, ,\\
		\Phi_\leftarrow^\mathrm{e} (x) &= \sqrt{\frac{\mathcal N_E}{\mathcal L}} \bigl(  0 , 0 , e^{\alpha/2} ,  e^{-\alpha/2} \bigr)^\mathrm{T} e^{-\i k x} \, \Theta(-x) \, , \\
		\Phi_\rightarrow^\mathrm{e} (x) &= \sqrt{\frac{\mathcal N_E}{\mathcal L}} \bigl( e^{\alpha/2} ,  e^{-\alpha/2} , 0 , 0 \bigr)^\mathrm{T} e^{\i k x} \, \Theta(x) \, ,\\
		\Phi_\rightarrow^\mathrm{h} (x) &= \sqrt{\frac{\mathcal N_E}{\mathcal L}} \bigl(  0 , 0 , e^{-\alpha/2} ,  e^{\alpha/2} \bigr)^\mathrm{T} e^{\i k x} \, \Theta(x) \, , 
	\end{align}
	\label{eqn:outgoing}
\end{subequations}
where $\Phi_q^r$ is a $r$-like (e: electron, h: hole) quasiparticle moving to the $q$ ($\leftarrow$: left, $\rightarrow$: right) lead. Here, we defined the energy dependent scattering phase $\alpha(E)$ via
\begin{align}
e^{\pm \alpha(E)} = \frac{E}{\Delta} \pm \sqrt{\frac{E^2}{\Delta^2} - 1} \, ,
\label{eqn:energyScattering2}
\end{align}
and the wave number 
\begin{equation}
	k(E) = \frac{\sqrt{E^2 - \Delta^2}}{\hbar v_\cj} \, .
\end{equation}
Moreover, $\mathcal N_E = \Delta/2E$ is a normalization constant and $\mathcal L$ is a length-scale over which the propagating quasiparticle states are defined.

In the same way, we define the four possible incident quasiparticles as
\begin{subequations}
	\begin{align}
	\Phi_1^\mathrm{in} (x) &= \sqrt{\frac{\mathcal N_E}{\mathcal L}} \bigl( e^{\alpha/2} ,  e^{-\alpha/2} , 0 , 0 \bigr)^\mathrm{T} e^{\i k x} \, \Theta(-x) \, ,\\
	\Phi_2^\mathrm{in} (x) &= \sqrt{\frac{\mathcal N_E}{\mathcal L}} \bigl(  0 , 0 , e^{-\alpha/2} ,  e^{\alpha/2} \bigr)^\mathrm{T} e^{\i k x} \, \Theta(-x) \, , \\
	\Phi_3^\mathrm{in} (x) &= \sqrt{\frac{\mathcal N_E}{\mathcal L}} \bigl( e^{-\alpha/2} ,  e^{\alpha/2} , 0 , 0 \bigr)^\mathrm{T} e^{-\i k x} \, \Theta(x) \, ,\\
	\Phi_4^\mathrm{in} (x) &= \sqrt{\frac{\mathcal N_E}{\mathcal L}} \bigl(  0 , 0 , e^{\alpha/2} ,  e^{-\alpha/2} \bigr)^\mathrm{T} e^{-\i k x} \, \Theta(x) \, .
	\end{align}
	\label{eqn:incident}
\end{subequations}
With the incident and outgoing quasiparticles defined in Eqs.\,(\ref{eqn:incident}) and (\ref{eqn:outgoing}), respectively, we write the four scattering states as
\begin{multline}
	\Phi_{E,s}(x) = \Phi_s^\mathrm{in}(x) +  A_s \Phi_\leftarrow^\mathrm{h}(x) +  B_s \Phi_\leftarrow^\mathrm{e}(x) 
	\\
	+ C_s \Phi_\rightarrow^\mathrm{e}(x) + D_s \Phi_\rightarrow^\mathrm{h}(x) \, .
	%
\label{eqn:scatteringStates}
\end{multline}
The corresponding scattering coefficients $A_s$, $B_s$, $C_s$ and $D_s$ are obtained by using the boundary condition in Eq.\,(\ref{eqn:boundary}), i.e. solving the equation
\begin{equation}
	\Phi_{E,s}(0^+) = B(0^+,0^-) \, \Phi_{E,s}(0^-)
\end{equation}
for each of the four scattering states. The full solution reads
\begin{subequations}
	\begin{align}
		\bigl(
			A_1 , B_1 , C_1 , D_1 
		\bigr)
		&=
		\bigl(
			\mathcal A , \mathcal B , \mathcal C , \mathcal D 
		\bigr)
		\, , \\
		\bigl(
			A_2 , B_2 , C_2 , D_2 
		\bigr)
		&=
		\bigl(
			\mathcal B^* , \mathcal A^* , \mathcal D^* , \mathcal C^*
		\bigr)
		\, , \\
		\bigl(
			A_3 , B_3 , C_3 , D_3
		\bigr)
		&=
		\bigl(
			-\mathcal D^* , \mathcal C^* , -\mathcal B^* , \mathcal A^*
		\bigr)
		\, , \\
		\bigl(
			A_4 , B_4 , C_4 , D_4
		\bigr)
		&=
		\bigl(
			\mathcal C , - \mathcal D , \mathcal A , - \mathcal B
		\bigr)
	\end{align}
\end{subequations}
with the scattering coefficients
\begin{subequations}
	\begin{align}
		\mathcal A &= \frac{\Delta^2}{E^2 - E_\mathrm{A}^2} \, (- \i \mathcal T) \, \sin\frac{\varphi}{2} \, \sinh\Bigl( \alpha - \i \frac{\varphi}{2} \Bigr) \, ,  \\
		\mathcal B &= \frac{\Delta^2}{E^2 - E_\mathrm{A}^2} \, (- \i \sqrt{1-\mathcal T}) \, \sinh^2\alpha \, , \\
		\mathcal C &= \frac{\Delta^2}{E^2 - E_\mathrm{A}^2} \, \sqrt{\mathcal T} \, \sinh\alpha \, \sinh\Bigl( \alpha - \i \frac{\varphi}{2} \Bigr) \, ,  \\
		\mathcal D &= \frac{\Delta^2}{E^2 - E_\mathrm{A}^2} \, \sqrt{1-\mathcal T} \, \sqrt{\mathcal T} \, \sin\frac{\varphi}{2}\, \sinh\alpha \, ,
	\end{align}
\end{subequations}
satisfying $\mathcal A \mathcal B + \mathcal C \mathcal D = 0$ and $|\mathcal A|^2 + |\mathcal B|^2 + |\mathcal C|^2 + |\mathcal D|^2 = 1$. Here, we already introduced $E_\mathrm{A} = \Delta \sqrt{1 - \mathcal T \sin^2(\varphi/2)}$ which is the energy of the Andreev bound state inside the gap (see appendix \ref{subappendix:a2b}).

\subsubsection{Andreev bound state wave function}
\label{subappendix:a2b}
For energies $E<\Delta$, there is no incident quasiparticle and $\alpha \to \i\alpha$ and $k \to \i k$ become complex. Therefore, the scattering state for the Andreev bound state following from Eq.\,(\ref{eqn:scatteringStates}) reads
\begin{multline}
	\Phi_\mathrm{A}(x) 
	= 
	\begin{pmatrix} e^{-\i \alpha / 2} A_0 \\ e^{\i \alpha / 2} A_0 \\ e^{\i \alpha / 2} B_0 \\ e^{-\i \alpha / 2} B_0 \end{pmatrix} e^{k x} \, \Theta(-x) 
	\\
	+ 
	\begin{pmatrix} e^{\i \alpha / 2} C_0 \\ e^{-\i \alpha / 2} C_0 \\ e^{-\i \alpha / 2} D_0 \\ e^{\i \alpha / 2} D_0 \end{pmatrix} e^{-k x} \, \Theta(x) \, ,
\end{multline}
with $\alpha(E)$ defined via
\begin{align}
e^{\pm \i \alpha(E)} = \frac{E}{\Delta} \pm \i \sqrt{ 1 - \frac{E^2}{\Delta^2} } \, .
\label{eqn:energyScattering2}
\end{align}
and the wave number
\begin{equation}
	k(E)  = \frac{\sqrt{\Delta^2 - E^2}}{\hbar v_\cj}  \, .
\end{equation}
Applying the boundary condition in Eq.\,(\ref{eqn:boundary}), i.e.
\begin{equation}
	\Phi_\mathrm{A}(0^+) = B(0^+,0^-) \, \Phi_\mathrm{A}(0^-) \, ,
	\label{eqn:boundaryAndreev}
\end{equation}
we find the energy has to fulfill $E = E_\mathrm{A}$ with the Andreev bound state energy
\begin{equation}
	E_\mathrm{A}(\varphi,\mathcal T) = \Delta \sqrt{1 - \mathcal T \sin^2\frac{\varphi}{2}} \, ,
	\label{eq:ABSenergy}
\end{equation}
in order to find nontrivial solutions for the scattering coefficients $A_0$, $B_0$, $C_0$ and $D_0$. Under this condition, Eq.\,(\ref{eqn:boundaryAndreev}) reveals
\begin{subequations}
	\begin{align}
		A_0 &= -C_0 \, , \\
		B_0 &= D_0 \, , \\
		B_0 &= -\frac{\i }{\sqrt{1-\mathcal T}} \left( \sqrt{\mathcal T} \cos \frac{\varphi}{2} + \frac{E_\mathrm{A}}{\Delta}\right) A_0 \, .
	\end{align}
\end{subequations}
This notation is valid for $\varphi \in [0 , \pi]$ and the case $\mathcal T = 1$ needs a special treatment since the ABS energy as defined in Eq.\,(\ref{eq:ABSenergy}) is $E_\mathrm{A} \geq 0$ for all phases. Signs have to be adjusted for phases $\varphi \in [\pi,2\pi]$.

To normalize the subgap wave function, we use
\begin{equation}
	1 = \int_{-\infty}^\infty \!\! \! \d x \, \bigl( \Phi_\mathrm{A}^*(x) \bigr)^\mathrm{T} \Phi_\mathrm{A}(x) = \frac{2}{k_\mathrm{A}} \Bigl( |A_0|^2 + |B_0|^2 \Bigr) \, ,
\end{equation}
with $k_\mathrm{A} = k(E_\mathrm{A})$, and finally obtain the solution
\begin{subequations}
	\begin{align}
		A_0 &= \sqrt{N_0} \sqrt{1-\mathcal T} \sin \frac{\varphi}{2} \, ,\\
		B_0 &= - \i \sqrt{N_0} \left( \sqrt{\mathcal T} \cos \frac{\varphi}{2} + \frac{E_\mathrm{A}}{\Delta}\right)  \sin \frac{\varphi}{2} \, , \\
		N_0 &= \frac{ \Delta^2 \sqrt{\mathcal T}  }{ 4 \hbar v_\cj E_\mathrm{A} \sin \frac{\varphi}{2} \left( \sqrt{\mathcal T} \cos \frac{\varphi}{2} + \frac{E_\mathrm{A}}{\Delta}\right)} \, ,
	\end{align}
\end{subequations}
satisfying $A_0 B_0 + C_0 D_0 = 0$.

%
%
%
\section{Current operator matrix elements}
\label{appendix:b}
In this appendix, we provide the matrix elements of the current operator for the short topological (appendix \ref{subappendix:b1}) and the short conventional (appendix \ref{subappendix:b2}) superconducting junction.

First, we are going to write the current operator of Eq.\,(\ref{eqn:currentOperatorSSJ}) in terms of the solutions of the Bogoliubov-de Gennes equation for the topological (conventional) junction obtained in appendix \ref{subappendix:a1} (appendix \ref{subappendix:a2}). Since the Hamiltonian $\mathcal H_{\tj(\cj)}(x)$ in Eq.\,(\ref{ch4:bdgHamiltonianTopologicalJunction}) (Eq.\,(\ref{ch4:bdgHamiltonianConventionalJunction})) obeys particle-hole symmetry, described in Sec.\,\ref{subsec:ssj}, we can write the spinors for each junction as
\begin{subequations}
	\begin{align}
		\Psi_{\tj}(0^-) &= \sum_{n>0} \Bigl\{ \Phi_n(0^-) \gamma_n^{\phantom{\dag}} + (\mathcal S_\tj \Phi_n(0^-) ) \gamma_n^\dag \Bigr\} \, , \label{eqnApp:spinorTJ}\\
		\Psi_{\cj}(0^-) &= \sum_{n>0} \Bigl\{ \Phi_n(0^-) \gamma_{1,n}^{\phantom{\dag}} + (\mathcal S_\cj \Phi_n(0^-) ) \gamma_{2,n}^\dag \Bigr\} \, , \label{eqnApp:spinorCJ}
	\end{align}
\end{subequations}
by using only the solutions at positive energy, with $n = (E,s)$ for continuum states $s \in \{1,2,3,4\}$ with energy $E>\Delta$ and $n = E_\mathrm{M(A)}$ for the Majorana (Andreev) bound state in the topological (conventional) junction. We note that quasiparticle states in the conventional junction are spin-degenerate. Therefore, particle-hole symmetry $\mathcal S_\cj = \i \sigma_1 \tau_2 K$ yields $\gamma_{1,-n}^{\phantom{\dag}} = \gamma_{2,n}^{\dag}$ with the definition of the spinor of spin-down quasiparticles $\mathcal S_\cj \Psi_{\cj}(x) = \bigl( \psi_{\R\downarrow}^{\phantom{\dag}}(x) , - \psi_{\L\uparrow}^{\dag }(x) , \psi_{\L\downarrow}^{\phantom{\dag}}(x) , - \psi_{\R\uparrow}^\dag(x)\bigr)^\text{T}$ denoted by the index "2".

\subsection{Matrix elements for the topological junction}
\label{subappendix:b1}
Using the definition of the spinor in Eq.\,(\ref{eqnApp:spinorTJ}), we calculate the current operator in Eq.\,(\ref{eqn:currentOperatorSSJ}) yielding
\begin{align}
	I_\tj &= \sum_{m,n > 0} I_{mn}^\mathrm{one} \bigl(2  \gamma_m^\dag \gamma_n^{\phantom{\dag}} - 1 \bigr) \nonumber \\
	&\quad + \sum_{m,n > 0} \bigl( I_{mn}^\mathrm{two} \,  \gamma_m^\dag \gamma_n^\dag + \hc \bigr) \, ,
\end{align}
with the matrix elements
\begin{subequations}
	\begin{align}
		I_{mn}^\mathrm{one} &= \frac{e v_\tj}{2} \Phi_m^{*\mathrm{T}}(0^-) \, \sigma_3 \,  \Phi_n^{\phantom{\mathrm{T}}}(0^-) \, , \\
		I_{mn}^\mathrm{two} &= \frac{e v_\tj}{2} \Phi_m^{*\mathrm{T}}(0^-) \, \sigma_3 \,  \mathcal S_\tj \Phi_n^{\phantom{\mathrm{T}}}(0^-) \, ,
	\end{align}
\end{subequations}
describing transitions of one or two quasiparticles, respectively. As already introduced in Sec.\,\ref{subsubsec:topological}, $\mathcal S_\tj =  \sigma_2 \tau_2 K$ describes particle-hole symmetry in the topological junction. Choosing $m=n=E_\mathrm{M}$, we obtain the supercurrent $I_\mathrm{M}$ carried by the MBS as
\begin{align}
	I_\mathrm{M} = - \frac{e \Delta}{\hbar} \, \sin\frac{\varphi}{2} \left( n_\mathrm{M} - \frac{1}{2} \right) = \frac{2e}{\hbar} \frac{\partial E_\mathrm{M}}{\partial \varphi} \left( n_\mathrm{M} - \frac{1}{2} \right) \, ,
	\label{eqn:subgapcurrentMBS}
\end{align}
with $n_\mathrm{M}$ being the occupation of the MBS. This is the supercurrent presented in Eq.\,(\ref{eqn:MBScurrent}) in the main text.

We note that the transfer of a Cooper pair from the ground state to the MBS is not possible because the corresponding matrix element $I_\mathrm{MM}^\mathrm{two} = 0$. Moreover, the matrix elements for transitions involving both the MBS and continuum states are given by
\begin{align}
	|I_{\mathrm{M}E}^\mathrm{one/two}|^2 &= \frac{e^2}{4 \pi \hbar^2 N_\tj}\frac{E^2 - \Delta^2}{ E} \frac{\sqrt{\Delta^2 - E_\mathrm{M}^2}}{E \mp E_\mathrm{M}}\, ,
\end{align}
where we defined $|I_{\mathrm{M}E}^\mathrm{one/two}|^2 = \sum_{s=1}^4 |I_{\mathrm{M}(E,s)}^\mathrm{one/two}|^2$ and introduced the density of states $N_\tj = \mathcal L/\pi\hbar v_\tj$ in one dimension in the normal state of the topological junction.

\subsection{Matrix elements for the conventional junction}
\label{subappendix:b2}
Using the definition of the spinor in Eq.\,(\ref{eqnApp:spinorCJ}), we calculate the current operator in Eq.\,(\ref{eqn:currentOperatorSSJ}) yielding
\begin{align}
	I_\cj &= \sum_{m,n > 0} I_{mn}^\mathrm{one} \bigl( \gamma_{1,m}^\dag \gamma_{1,n}^{\phantom{\dag}}  + \gamma_{2,m}^\dag \gamma_{2,n}^{\phantom{\dag}}  - 1 \bigr) 
	\nonumber \\
	& \quad + \sum_{m,n > 0} \bigl( I_{mn}^\mathrm{two} \,  \gamma_{1,m}^\dag \gamma_{2,n}^\dag + \hc \bigr) \, ,
\end{align}
with the matrix elements
\begin{subequations}
	\begin{align}
	I_{mn}^\mathrm{one} &= \frac{e v_\cj}{2} \Phi_m^{*\mathrm{T}}(0^-) \, \sigma_3 \,  \Phi_n^{\phantom{\mathrm{T}}}(0^-) \, , \\
	I_{mn}^\mathrm{two} &= \frac{e v_\cj}{2} \Phi_m^{*\mathrm{T}}(0^-) \, \sigma_3 \,  \mathcal S_\cj \Phi_n^{\phantom{\mathrm{T}}}(0^-) \, ,
	\end{align}
\end{subequations}
describing transitions of one or two quasiparticles, respectively. As already introduced in Sec.\,\ref{subsubsec:conventional}, $\mathcal S_\cj =  \i \sigma_1 \tau_2 K$ describes particle-hole symmetry in the conventional junction. Choosing $m=n=E_\mathrm{A}$, we obtain the supercurrent $I_\mathrm{A}$ carried by the Andreev bound state (ABS) as
\begin{align}
I_\mathrm{A} = - \frac{e \Delta^2}{4\hbar} \, \mathcal T \, \frac{\sin\varphi}{E_\mathrm{A}} \bigl( n_\mathrm{A} - 1 \bigr) = \frac{1}{2} \frac{2e}{\hbar} \, \frac{\partial E_\mathrm{A}}{\partial \varphi} \left( n_\mathrm{A} - 1 \right)  \, ,
\label{eqn:subgapcurrentABS}
\end{align}
with $n_\mathrm{A} = n_\mathrm{1,A} + n_\mathrm{2,A}$ being the occupation of the twofold degenerate ABS. The factor 1/2 is a result of the fact that the Bogoliubov-de Gennes Hamiltonian of the conventional junction describes only spin-up particles, while spin-down particles give the same contribution. The supercurrent presented in Eq.\,(\ref{eqn:ABScurrent}) in the main text is the full current $2 I_\mathrm{A}$ taking both spins into account.

The matrix element describing the transition of a Cooper pair between the ABS and the ground state is given by 
\begin{align}
	|I_\mathrm{AA}^\mathrm{two}|^2 = \frac{e^2}{4\hbar^2} (1-\mathcal T)\, \frac{(\Delta^2 - E_\mathrm{A}^2)^2}{E_\mathrm{A}^2} \, .
\end{align}
Moreover, the matrix elements for transitions involving both the ABS and continuum states are given by
\begin{multline}
|I_{\mathrm{A}E}^\mathrm{one/two}|^2 = \frac{e^2 \mathcal T}{4\pi \hbar^2 N_\cj } \, \frac{\sqrt{\Delta^2 - E_\mathrm{A}^2}}{E_\mathrm{A} \, E} \, \frac{E^2 - \Delta^2}{E^2 - E_\mathrm{A}^2}
\\
\times \Bigl( E_\mathrm{A} (E \mp E_\mathrm{A}) \pm \Delta^2 (\cos \varphi + 1) \Bigr)
\end{multline}
where we defined $|I_{\mathrm{A}E}^\mathrm{one/two}|^2 = \sum_{s=1}^4 |I_{\mathrm{A}(E,s)}^\mathrm{one/two}|^2$ and introduced the density of states $N_\cj = \mathcal L/\pi\hbar v_\cj$ in one dimension in the normal state of the conventional junction.

\bibliography{references}

\end{document}